\input harvmac
\noblackbox

\input epsf

\newcount\figno
\figno=0
\def\fig#1#2#3{
\par\begingroup\parindent=0pt\leftskip=1cm\rightskip=1cm\parindent=0pt
\baselineskip=11pt
\global\advance\figno by 1
\midinsert
\epsfxsize=#3
\centerline{\epsfbox{#2}}
\vskip 12pt
{\bf Fig.\ \the\figno: } #1\par
\endinsert\endgroup\par
}
\def\figlabel#1{\xdef#1{\the\figno}}
\def\encadremath#1{\vbox{\hrule\hbox{\vrule\kern8pt\vbox{\kern8pt
\hbox{$\displaystyle #1$}\kern8pt}
\kern8pt\vrule}\hrule}}

\font\cmss=cmss10
\font\cmsss=cmss10 at 7pt
\def\rlx{\relax\leavevmode}
\def\inbar{\vrule height1.5ex width.4pt depth0pt}
\def\IN{\relax{\rm I\kern-.18em N}}
\def\IP{\relax{\rm I\kern-.18em P}}
\def\ZZ{\rlx\leavevmode\ifmmode\mathchoice{\hbox{\cmss Z\kern-.4em Z}}
 {\hbox{\cmss Z\kern-.4em Z}}{\lower.9pt\hbox{\cmsss Z\kern-.36em Z}}
 {\lower1.2pt\hbox{\cmsss Z\kern-.36em Z}}\else{\cmss Z\kern-.4em
 Z}\fi}
\def\IZ{\relax\ifmmode\mathchoice
{\hbox{\cmss Z\kern-.4em Z}}{\hbox{\cmss Z\kern-.4em Z}}
{\lower.9pt\hbox{\cmsss Z\kern-.4em Z}}
{\lower1.2pt\hbox{\cmsss Z\kern-.4em Z}}\else{\cmss Z\kern-.4em
Z}\fi}
\def\IZ{\relax\ifmmode\mathchoice
{\hbox{\cmss Z\kern-.4em Z}}{\hbox{\cmss Z\kern-.4em Z}}
{\lower.9pt\hbox{\cmsss Z\kern-.4em Z}}
{\lower1.2pt\hbox{\cmsss Z\kern-.4em Z}}\else{\cmss Z\kern-.4em
Z}\fi}

\def\narrowplus{\kern -.04truein + \kern -.03truein}
\def\narrowminus{- \kern -.04truein}
\def\narrowminussub{\kern -.02truein - \kern -.01truein}

\def\th{{\theta}}

\def\s{{\sigma}}
\def\frac#1#2{{#1\over #2}}

\def\IZ{\relax\ifmmode\mathchoice
{\hbox{\cmss Z\kern-.4em Z}}{\hbox{\cmss Z\kern-.4em Z}}
{\lower.9pt\hbox{\cmsss Z\kern-.4em Z}}
{\lower1.2pt\hbox{\cmsss Z\kern-.4em Z}}\else{\cmss Z\kern-.4em
Z}\fi}
\def\IC{{\relax\,\hbox{$\inbar\kern-.3em{\rm C}$}}}

\font\cmss=cmss10 \font\cmsss=cmss10 at 7pt
\def\IR{\relax{\rm I\kern-.18em R}}
\def\ra{\rangle}
\def\la{\langle}

\def\IZ{\relax\ifmmode\mathchoice
{\hbox{\cmss Z\kern-.4em Z}}{\hbox{\cmss Z\kern-.4em Z}}
{\lower.9pt\hbox{\cmsss Z\kern-.4em Z}}
{\lower1.2pt\hbox{\cmsss Z\kern-.4em Z}}\else{\cmss Z\kern-.4em
Z}\fi}
\def\IC{{\relax\,\hbox{$\inbar\kern-.3em{\rm C}$}}}
\def\O{{\Omega}}

\def\a{{\alpha}}

\def\D{{\Delta}}
\def\m{{\mu}}

\def\ep{{\epsilon}}

\def\d{{\delta}}
\def\o{{\omega}}
\def\G{{\Gamma}}

\def\th{{\theta}}

\def\t{{\tau}}

\def\th{{\theta}}

\def\s{{\sigma}}

\def\R{{\rightarrow}}

\def\frac#1#2{{#1\over #2}}

\def\L{{\Lambda}}

\def\tt{{\tilde{\tau}}}
\def\tz{\tilde{z}}
\def\mt{|t|}

\lref\ffstr{R.~Gopakumar, ``Free field theory as a string theory?,''
  Comptes Rendus Physique {\bf 5}, 1111 (2004)
  [arXiv:hep-th/0409233].}
\lref\ffa{R.~Gopakumar,
``From free fields to AdS,''
Phys.\ Rev.\ D {\bf 70}, 025009 (2004)
  [arXiv:hep-th/0308184].}
\lref\ffb{R.~Gopakumar,
``From free fields to AdS - II,''
Phys.\ Rev.\ D {\bf 70}, 025010 (2004)
  [arXiv:hep-th/0402063].}
\lref\thft{G.~'tHooft, 
``A Planar Diagram Theory for Strong Interactions,'' Nucl.\ Phys.\
{\bf 72}, 461, (1974).}
\lref\gv{
R.~Gopakumar and C.~Vafa,
``On the gauge theory/geometry correspondence,''
Adv.\ Theor.\ Math.\ Phys.\  {\bf 3}, 1415 (1999)
[arXiv:hep-th/9811131].}
\lref\ov{
H.~Ooguri and C.~Vafa,
``Worldsheet derivation of a large N duality,''
Nucl.\ Phys.\ B {\bf 641}, 3 (2002)
[arXiv:hep-th/0205297].}
\lref\akh{E.~T.~Akhmedov,
  ``Expansion in Feynman graphs as simplicial string theory,''
  JETP Lett.\  {\bf 80}, 218 (2004)
  [Pisma Zh.\ Eksp.\ Teor.\ Fiz.\  {\bf 80}, 247 (2004)]
  [arXiv:hep-th/0407018].}
\lref\car{M.~Carfora, C.~Dappiaggi and V.~Gili,
  ``Simplicial aspects of string dualities,''
  AIP Conf.\ Proc.\  {\bf 751}, 182 (2005)
  [arXiv:hep-th/0410006].}
\lref\gor{A.~Gorsky and V.~Lysov,
  ``From effective actions to the background geometry,''
  [arXiv:hep-th/0411063].}
\lref\justin{
L.~F.~Alday, J.~R.~David, E.~Gava and K.~S.~Narain,
  ``Structure constants of planar N = 4 Yang Mills at one loop,''
  arXiv:hep-th/0502186.}
\lref\kont{M.~Kontsevich,
  ``Intersection theory on the moduli space of curves and the matrix Airy
  function,''
  Commun.\ Math.\ Phys.\  {\bf 147}, 1 (1992).}
\lref\harer{J. Harer, ``The cohomology of the moduli space of curves,''
Lecture Notes in Math. vol. {\bf 1337} pp. 138-221.}
\lref\pen{R. ~C. ~Penner, The Decorated Teichmuller theory of punctured 
surfaces,''Commun.\ Math.\ Phys.\  {\bf 113}, 299 (1987).}
\lref\witint{E.~Witten,
  ``Two-Dimensional Gravity And Intersection Theory On Moduli Space,''
  Surveys Diff.\ Geom.\  {\bf 1}, 243 (1991).}
\lref\streb{K. ~Strebel, ``Quadratic Differentials,'' Springer-Verlag
(1980).}

\lref\bov{N.~Berkovits, H.~Ooguri and C.~Vafa,
``On the worldsheet derivation of large N dualities for the superstring,''
arXiv:hep-th/0310118.}
\lref\malda{
J.~M.~Maldacena,
``The large $N$ limit of superconformal field theories and supergravity,''
Adv.\ Theor.\ Math.\ Phys.\  {\bf 2}, 231 (1998)
[Int.\ J.\ Theor.\ Phys.\  {\bf 38}, 1113 (1999)]
[arXiv:hep-th/9711200].}

\lref\divec{P.~Di Vecchia, L.~Magnea, A.~Lerda, R.~Russo and R.~Marotta,
``String techniques for the calculation of renormalization constants in field theory,''
Nucl.\ Phys.\ B {\bf 469}, 235 (1996)
[arXiv:hep-th/9601143].}
\lref\fmr{A.~Frizzo, L.~Magnea and R.~Russo,
``Systematics of one-loop Yang-Mills diagrams from bosonic string  amplitudes,''
Nucl.\ Phys.\ B {\bf 604}, 92 (2001)
[arXiv:hep-ph/0012129].}
\lref\berkos{
Z.~Bern and D.~A.~Kosower,
``Efficient Calculation Of One Loop QCD Amplitudes,''
Phys.\ Rev.\ Lett.\  {\bf 66}, 1669 (1991).}
\lref\bdh{
L.~Brink, P.~Di Vecchia and P.~S.~Howe,
``A Lagrangian Formulation Of The Classical And Quantum Dynamics Of Spinning Particles,''
Nucl.\ Phys.\ B {\bf 118}, 76 (1977).}
\lref\ssen{M.~Tuite and S.~Sen,
``A String Motivated Approach to the Relativistic Point Particle,''
arXiv:hep-th/0308099.}
\lref\strass{M.~J.~Strassler,
``Field theory without Feynman diagrams: One loop effective actions,''
Nucl.\ Phys.\ B {\bf 385}, 145 (1992)
[arXiv:hep-ph/9205205].}
\lref\schb{C.~Schubert,
``Perturbative quantum field theory in the string-inspired formalism,''
Phys.\ Rept.\  {\bf 355}, 73 (2001)
[arXiv:hep-th/0101036].}
\lref\mikh{A.~Mikhailov,
``Notes on higher spin symmetries,''
arXiv:hep-th/0201019.}
\lref\cnss{G.~Chalmers, H.~Nastase, K.~Schalm and R.~Siebelink,
``R-current correlators in N = 4 super Yang-Mills theory from anti-de  Sitter supergravity,''
Nucl.\ Phys.\ B {\bf 540}, 247 (1999)
[arXiv:hep-th/9805105].}
\lref\guil{E. A. Guillemin, ``Introductory Circuit Theory'', (John Wiley and Sons, 1953).}
\lref\BD{J. D. Bjorken and S. D. Drell, ``Relativistic Quantum Fields'', (McGraw Hill, 1965).}
\lref\iz{C.~Itzykson and J-B.~Zuber, ``Quantum Field Theory,'' (Mc Graw Hill, (1980).}
\lref\lam{C.~S.~Lam,
``Multiloop string - like formulas for QED,''
Phys.\ Rev.\ D {\bf 48}, 873 (1993)
[arXiv:hep-ph/9212296].}
\lref\lama{C.~S.~Lam,
``Spinor helicity technique and string reorganization for multiloop
diagrams,''
Can.\ J.\ Phys.\  {\bf 72}, 415 (1994)
[arXiv:hep-ph/9308289].}
\lref\petkou{A.~Petkou,
``Conserved currents, consistency relations, and operator product  expansions
in the conformally invariant O(N) vector model,''
Annals Phys.\  {\bf 249}, 180 (1996)
[arXiv:hep-th/9410093].}
\lref\liu{H.~Liu,
``Scattering in anti-de Sitter space and operator product expansion,''
Phys.\ Rev.\ D {\bf 60}, 106005 (1999)
[arXiv:hep-th/9811152].}
\lref\dmmr{E.~D'Hoker, S.~D.~Mathur, A.~Matusis and L.~Rastelli,
``The operator product expansion of N = 4 SYM and the 4-point functions  of supergravity,''
Nucl.\ Phys.\ B {\bf 589}, 38 (2000)
[arXiv:hep-th/9911222].}
\lref\dfmmr{E.~D'Hoker, D.~Z.~Freedman, S.~D.~Mathur, A.~Matusis and L.~Rastelli,
``Graviton exchange and complete 4-point functions in the AdS/CFT  correspondence,''
Nucl.\ Phys.\ B {\bf 562}, 353 (1999)
[arXiv:hep-th/9903196].}
\lref\fmmr{D.~Z.~Freedman, S.~D.~Mathur, A.~Matusis and L.~Rastelli,
``Correlation functions in the CFT($d$)/AdS($d+1$) correspondence,''
Nucl.\ Phys.\ B {\bf 546}, 96 (1999)
[arXiv:hep-th/9804058].}
\lref\sunda{B.~Sundborg,
``The Hagedorn transition, deconfinement and N = 4 SYM theory,''
Nucl.\ Phys.\ B {\bf 573}, 349 (2000)
[arXiv:hep-th/9908001].}
\lref\sund{P.~Haggi-Mani and B.~Sundborg,
``Free large N supersymmetric Yang-Mills theory as a string theory,''
JHEP {\bf 0004}, 031 (2000)
[arXiv:hep-th/0002189];}
\lref\sundb{B.~Sundborg,
``Stringy gravity, interacting tensionless strings and massless higher  spins,''
Nucl.\ Phys.\ Proc.\ Suppl.\  {\bf 102}, 113 (2001)
[arXiv:hep-th/0103247].}
\lref\witt{E. Witten, Talk at the Schwarzfest,
http://theory.caltech.edu/jhs60/witten.}
\lref\tset{A.~A.~Tseytlin,
``On limits of superstring in AdS(5) x S**5,''
Theor.\ Math.\ Phys.\  {\bf 133}, 1376 (2002)
[Teor.\ Mat.\ Fiz.\  {\bf 133}, 69 (2002)]
[arXiv:hep-th/0201112].}
\lref\mz{J.~A.~Minahan and K.~Zarembo,
``The Bethe-ansatz for N = 4 super Yang-Mills,''
JHEP {\bf 0303}, 013 (2003)
[arXiv:hep-th/0212208].}
\lref\dmw{A.~Dhar, G.~Mandal and S.~R.~Wadia,
``String bits in small radius AdS and weakly coupled N = 4 super  Yang-Mills theory. I,''
arXiv:hep-th/0304062.}
\lref\dnw{L.~Dolan, C.~R.~Nappi and E.~Witten,
``A relation between approaches to integrability in superconformal Yang-Mills theory,''
arXiv:hep-th/0308089.}
\lref\polch{J.~Polchinski, Seminar at SW workshop on String Theory (Feb. 2003), unpublished.}
\lref\msw{G.~Mandal, N.~V.~Suryanarayana and S.~R.~Wadia,
``Aspects of semiclassical strings in AdS(5),''
Phys.\ Lett.\ B {\bf 543}, 81 (2002)
[arXiv:hep-th/0206103].}
\lref\bnp{I.~Bena, J.~Polchinski and R.~Roiban,
``Hidden symmetries of the AdS(5) x S**5 superstring,''
arXiv:hep-th/0305116.}
\lref\vall{B.~C.~Vallilo,
``Flat currents in the classical AdS(5) x S**5 pure spinor superstring,''
arXiv:hep-th/0307018.}
\lref\klepol{I.~R.~Klebanov and A.~M.~Polyakov,
``AdS dual of the critical O(N) vector model,''
Phys.\ Lett.\ B {\bf 550}, 213 (2002)
[arXiv:hep-th/0210114].}
\lref\peta{A.~C.~Petkou,
``Evaluating the AdS dual of the critical O(N) vector model,''
JHEP {\bf 0303}, 049 (2003)
[arXiv:hep-th/0302063].}
\lref\leigh{R.~G.~Leigh and A.~C.~Petkou,
``Holography of the N = 1 higher-spin theory on AdS(4),''
JHEP {\bf 0306}, 011 (2003)
[arXiv:hep-th/0304217].}
\lref\sezsuna{E.~Sezgin and P.~Sundell,
``Holography in 4D (super) higher spin theories and a test via cubic  scalar couplings,''
arXiv:hep-th/0305040.}
\lref\por{L.~Girardello, M.~Porrati and A.~Zaffaroni,
``3-D interacting CFTs and generalized Higgs phenomenon in higher spin  theories on AdS,''
Phys.\ Lett.\ B {\bf 561}, 289 (2003)
[arXiv:hep-th/0212181].}
\lref\ruhl{T.~Leonhardt, A.~Meziane and W.~Ruhl,
``On the proposed AdS dual of the critical O(N) sigma model for any  dimension $2 < d < 4$,''
Phys.\ Lett.\ B {\bf 555}, 271 (2003)
[arXiv:hep-th/0211092].}
\lref\rajan{F.~Kristiansson and P.~Rajan,
``Scalar field corrections to AdS(4) gravity from higher spin gauge  theory,''
JHEP {\bf 0304}, 009 (2003)
[arXiv:hep-th/0303202].}
\lref\sumit{S.~R.~Das and A.~Jevicki,
``Large-N collective fields and holography,''
Phys.\ Rev.\ D {\bf 68}, 044011 (2003)
[arXiv:hep-th/0304093].}
\lref\nemani{N.~V.~Suryanarayana,
JHEP {\bf 0306}, 036 (2003)
[arXiv:hep-th/0304208].}
\lref\pol{A.~M.~Polyakov,
``Gauge fields and space-time,''
Int.\ J.\ Mod.\ Phys.\ A {\bf 17S1}, 119 (2002)
[arXiv:hep-th/0110196].}
\lref\polb{A.~M.~Polyakov,
``String theory and quark confinement,''
Nucl.\ Phys.\ Proc.\ Suppl.\  {\bf 68}, 1 (1998)
[arXiv:hep-th/9711002].}
\lref\polc{A.~M.~Polyakov,
``The wall of the cave,''
Int.\ J.\ Mod.\ Phys.\ A {\bf 14}, 645 (1999)
[arXiv:hep-th/9809057].}
\lref\polbk{A.~M.~Polyakov,
``Gauge fields and Strings,'' (Harwood Academic Publishers, 1987).}
\lref\bmn{D.~Berenstein, J.~M.~Maldacena and H.~Nastase,
``Strings in flat space and pp waves from N = 4 super Yang Mills,''
JHEP {\bf 0204}, 013 (2002)
[arXiv:hep-th/0202021].}
\lref\gkp{S.~S.~Gubser, I.~R.~Klebanov and A.~M.~Polyakov,
``Gauge theory correlators from non-critical string theory,''
Phys.\ Lett.\ B {\bf 428}, 105 (1998)
[arXiv:hep-th/9802109].}
\lref\witads{E.~Witten,
``Anti-de Sitter space and holography,''
Adv.\ Theor.\ Math.\ Phys.\  {\bf 2}, 253 (1998)
[arXiv:hep-th/9802150].}
\lref\wittwist{E.~Witten,
``Perturbative gauge theory as a string theory in twistor space,''
arXiv:hep-th/0312171.}
\lref\berktwist{N.~Berkovits,
``An Alternative String Theory in Twistor Space for N=4 Super-Yang-Mills,''
arXiv:hep-th/0402045.}
\lref\rsv{R.~Roiban, M.~Spradlin and A.~Volovich,
``A googly amplitude from the B-model in twistor space,''
arXiv:hep-th/0402016.}
\lref\sezsun{E.~Sezgin and P.~Sundell,
``Doubletons and 5D higher spin gauge theory,''
JHEP {\bf 0109}, 036 (2001)
[arXiv:hep-th/0105001]; ``Massless higher spins and holography,''
Nucl.\ Phys.\ B {\bf 644}, 303 (2002)
[Erratum-ibid.\ B {\bf 660}, 403 (2003)]
[arXiv:hep-th/0205131].}
\lref\vas{
M.~A.~Vasiliev,
``Conformal higher spin symmetries of 4D massless supermultiplets and  osp(L,2M)
invariant equations in generalized (super)space,''
Phys.\ Rev.\ D {\bf 66}, 066006 (2002)
[arXiv:hep-th/0106149].}
\lref\vasrev{M.~A.~Vasiliev,
``Higher spin gauge theories: Star-product and AdS space,''
arXiv:hep-th/9910096.}
\lref\bv{A.~O.~Barvinsky and G.~A.~Vilkovisky,
``Beyond The Schwinger-Dewitt Technique: Converting Loops Into Trees And In-In Currents,''
Nucl.\ Phys.\ B {\bf 282}, 163 (1987).}
\lref\birdav{N.~D.~Birrell and P.~C.~W.~Davies, ``Quantum fields in Curved Space,'' Cambridge University
Press (1984).}
\lref\hsken{M.~Henningson and K.~Skenderis,
``The holographic Weyl anomaly,''
JHEP {\bf 9807}, 023 (1998)
[arXiv:hep-th/9806087].}
\lref\nojod{S.~Nojiri and S.~D.~Odintsov,
``Conformal anomaly for dilaton coupled theories from AdS/CFT  correspondence,''
Phys.\ Lett.\ B {\bf 444}, 92 (1998)
[arXiv:hep-th/9810008].}
\lref\odnoj{S.~Nojiri, S.~D.~Odintsov and S.~Ogushi,
``Finite action in d5 gauged supergravity and dilatonic conformal anomaly  for dual quantum field theory,''
Phys.\ Rev.\ D {\bf 62}, 124002 (2000)
[arXiv:hep-th/0001122].}
\lref\dss{S.~de Haro, S.~N.~Solodukhin and K.~Skenderis,
``Holographic reconstruction of spacetime and renormalization in the  AdS/CFT correspondence,''
Commun.\ Math.\ Phys.\  {\bf 217}, 595 (2001)
[arXiv:hep-th/0002230].}
\lref\feffgr{C.~Fefferman, C.~R.~Graham, ``Conformal Invariants'' in ``{\it Elie Cartan et les
Mathematiques d'Aujourd'hui},'' (Asterisque 1985), 95.}
\lref\glebfrpet{G.~Arutyunov, S.~Frolov and A.~C.~Petkou,
``Operator product expansion of the lowest weight CPOs in N = 4  SYM(4) at strong coupling,''
Nucl.\ Phys.\ B {\bf 586}, 547 (2000)
[Erratum-ibid.\ B {\bf 609}, 539 (2001)]
[arXiv:hep-th/0005182];
``Perturbative and instanton corrections to the OPE of CPOs in N = 4  SYM(4),''
Nucl.\ Phys.\ B {\bf 602}, 238 (2001)
[Erratum-ibid.\ B {\bf 609}, 540 (2001)]
[arXiv:hep-th/0010137].}
\lref\arutfrol{G.~Arutyunov and S.~Frolov,
``Three-point Green function of the stress-energy tensor in the AdS/CFT  correspondence,''
Phys.\ Rev.\ D {\bf 60}, 026004 (1999)
[arXiv:hep-th/9901121].}
\lref\east{M.~G.~Eastwood,
``Higher symmetries of the Laplacian,''
arXiv:hep-th/0206233.}
\lref\gm{D.~J.~Gross and P.~F.~Mende,
``String Theory Beyond The Planck Scale,''
Nucl.\ Phys.\ B {\bf 303}, 407 (1988).}
\lref\gross{D.~J.~Gross,
``High-Energy Symmetries Of String Theory,''
Phys.\ Rev.\ Lett.\  {\bf 60}, 1229 (1988).}
\lref\bianchi{M.~Bianchi, J.~F.~Morales and H.~Samtleben,
``On stringy AdS(5) x S**5 and higher spin holography,''
JHEP {\bf 0307}, 062 (2003).}
\lref\bbms{N.~Beisert, M.~Bianchi, J.~F.~Morales and H.~Samtleben,
``On the spectrum of AdS/CFT beyond supergravity,''
arXiv:hep-th/0310292.}
\lref\anselmi{D.~Anselmi,
``The N = 4 quantum conformal algebra,''
Nucl.\ Phys.\ B {\bf 541}, 369 (1999)
[arXiv:hep-th/9809192].}
\lref\lind{U.~Lindstrom and M.~Zabzine,
``Tensionless strings, WZW models at critical level and massless higher  spin fields,''
arXiv:hep-th/0305098.}
\lref\witcube{E.~Witten,
``Noncommutative Geometry And String Field Theory,''
Nucl.\ Phys.\ B {\bf 268}, 253 (1986).}
\lref\gmw{S.~B.~Giddings, E.~J.~Martinec and E.~Witten,
``Modular Invariance In String Field Theory,''
Phys.\ Lett.\ B {\bf 176}, 362 (1986).}
\lref\zwie{B.~Zwiebach,
``A Proof That Witten's Open String Theory Gives A Single Cover Of Moduli
Space,''
Commun.\ Math.\ Phys.\  {\bf 142}, 193 (1991).}
\lref\penn{R.~Penner, ``Perturbative Series and the Moduli Space of
Riemann Surfaces,'' J. \ Diff.\ Geom.{\bf 27}, 35 (1988).}
\lref\sunil{S.~Mukhi,
``Topological matrix models, Liouville matrix model and c = 1 string theory,''
arXiv:hep-th/0310287.}
\lref\ashoke{A.~Sen, ``Open-closed duality at tree level,''
arXiv:hep-th/0306137.} 
\lref\gir{D.~Gaiotto, N.~Itzhaki and
L.~Rastelli, ``Closed strings as imaginary D-branes,''
arXiv:hep-th/0304192.} 
\lref\grsz{D.~Gaiotto, L.~Rastelli, A.~Sen and B.~Zwiebach,
``Ghost structure and closed strings in vacuum string field theory,''
Adv.\ Theor.\ Math.\ Phys.\  {\bf 6}, 403 (2003)
[arXiv:hep-th/0111129].}
\lref\gr{D.~Gaiotto and L.~Rastelli,
 ``A paradigm of open/closed duality: Liouville D-branes and the Kontsevich
model,''
arXiv:hep-th/0312196.}
\lref\satchi{S.~R.~Das, S.~Naik and S.~R.~Wadia, ``Quantization Of
The Liouville Mode And String Theory,'' Mod.\ Phys.\ Lett.\ A {\bf
4}, 1033 (1989).} 
\lref\spenta{A.~Dhar, T.~Jayaraman, K.~S.~Narain
and S.~R.~Wadia, ``The Role Of Quantized Two-Dimensional Gravity
In String Theory,'' Mod.\ Phys.\ Lett.\ A {\bf 5}, 863 (1990).}
\lref\karch{A.~Clark, A.~Karch, P.~Kovtun and D.~Yamada,
``Construction of bosonic string theory on infinitely curved
anti-de  Sitter space,'' arXiv:hep-th/0304107; A.~Karch,
``Lightcone quantization of string theory duals of free field
theories,'' arXiv:hep-th/0212041.} 
\lref\lmrs{S.~M.~Lee,
S.~Minwalla, M.~Rangamani and N.~Seiberg, ``Three-point functions
of chiral operators in D = 4, N = 4 SYM at  large N,'' Adv.\
Theor.\ Math.\ Phys.\  {\bf 2}, 697 (1998)
[arXiv:hep-th/9806074].} 
\lref\hsw{P.~S.~Howe, E.~Sokatchev and P.~C.~West,
``3-point functions in N = 4 Yang-Mills,''
Phys.\ Lett.\ B {\bf 444}, 341 (1998)
[arXiv:hep-th/9808162].}
\lref\dfs{E.~D'Hoker, D.~Z.~Freedman and W.~Skiba,
``Field theory tests for correlators in the AdS/CFT correspondence,''
Phys.\ Rev.\ D {\bf 59}, 045008 (1999)
[arXiv:hep-th/9807098].}
\lref\mettse{R.~R.~Metsaev and
A.~A.~Tseytlin, ``Type IIB superstring action in AdS(5) x S(5)
background,'' Nucl.\ Phys.\ B {\bf 533}, 109 (1998)
[arXiv:hep-th/9805028].} 
\lref\berk{N.~Berkovits, ``Super-Poincare
covariant quantization of the superstring,'' JHEP {\bf 0004}, 018
(2000) [arXiv:hep-th/0001035].} 
\lref\polchi{J.~Polchinski, Nucl.\
Phys.\ B {\bf 331}, 123, (1989).}
\lref\ymtherm{O.~Aharony, J.~Marsano, S.~Minwalla, K.~Papadodimas 
and M.~Van Raamsdonk,
``The Hagedorn / deconfinement phase transition in weakly coupled large N
gauge theories,''
arXiv:hep-th/0310285.}
\lref\thorn{K.~Bardakci and C.~B.~Thorn,
Nucl.\ Phys.\ B {\bf 626}, 287 (2002)
[arXiv:hep-th/0110301]; C.~B.~Thorn,
Nucl.\ Phys.\ B {\bf 637}, 272 (2002)
[Erratum-ibid.\ B {\bf 648}, 457 (2003)]
[arXiv:hep-th/0203167]; K.~Bardakci and C.~B.~Thorn,
Nucl.\ Phys.\ B {\bf 652}, 196 (2003)
[arXiv:hep-th/0206205]; S.~Gudmundsson, C.~B.~Thorn and T.~A.~Tran,
Nucl.\ Phys.\ B {\bf 649}, 3 (2003)
[arXiv:hep-th/0209102]; K.~Bardakci and C.~B.~Thorn,
Nucl.\ Phys.\ B {\bf 661}, 235 (2003)
[arXiv:hep-th/0212254]; C.~B.~Thorn and T.~A.~Tran,
Nucl.\ Phys.\ B {\bf 677}, 289 (2004)
[arXiv:hep-th/0307203].}
\lref\kumar{P.~de Medeiros and S.~P.~Kumar,
``Spacetime Virasoro algebra from strings on zero radius AdS(3),''
JHEP {\bf 0312}, 043 (2003)
[arXiv:hep-th/0310040].}
\lref\tse{A.~A.~Tseytlin,
``On semiclassical approximation and spinning string vertex operators in
AdS(5) x S**5,''
Nucl.\ Phys.\ B {\bf 664}, 247 (2003)
[arXiv:hep-th/0304139].}
\lref\mp{M. ~Mulase, M.~Penkava, ``Ribbon Graphs, 
Quadratic Differentials on Riemann Surfaces, and Algebraic 
Curves Defined over $\bar Q$,'' [math-ph/9811024].}
\lref\zvon{D. ~Zvonkine, ``Strebel differentials on stable curves and 
Kontsevich's proof of Witten's conjecture,'' [math.AG/0209071].}
\lref\seib{J.~Maldacena, G.~W.~Moore, N.~Seiberg and D.~Shih,
  ``Exact vs. semiclassical target space of the minimal string,''
  JHEP {\bf 0410}, 020 (2004)
  [arXiv:hep-th/0408039].}
\lref\giusto{S.~Giusto and C.~Imbimbo,
  Nucl.\ Phys.\ B {\bf 704}, 181 (2005)
  [arXiv:hep-th/0408216].}
\lref\aki{A.~Hashimoto, M.~x.~Huang, A.~Klemm and D.~Shih,
  ``Open / closed string duality for topological gravity with matter,''
  arXiv:hep-th/0501141.}
\lref\belzw{A.~Belopolsky and B.~Zwiebach,
  ``Off-shell closed string amplitudes: Towards a computation of the tachyon
  potential,''
  Nucl.\ Phys.\ B {\bf 442}, 494 (1995)
  [arXiv:hep-th/9409015].}
\lref\bel{A.~Belopolsky,
  ``Effective Tachyonic potential in closed string field theory,''
  Nucl.\ Phys.\ B {\bf 448}, 245 (1995)
  [arXiv:hep-th/9412106].}
\lref\moel{N.~Moeller,
  ``Closed bosonic string field theory at quartic order,''
  JHEP {\bf 0411}, 018 (2004)
  [arXiv:hep-th/0408067].}
\lref\yangz{H.~t.~Yang and B.~Zwiebach,
  ``Testing closed string field theory with marginal fields,''
  arXiv:hep-th/0501142.}
\lref\saadi{M.~Saadi and B.~Zwiebach,
  ``Closed String Field Theory From Polyhedra,''
  Annals Phys.\  {\bf 192}, 213 (1989).}
\lref\kugo{
T.~Kugo, H.~Kunitomo and K.~Suehiro,
  ``Nonpolynomial Closed String Field Theory,''
  Phys.\ Lett.\ B {\bf 226}, 48 (1989).}
\lref\wittop{E.~Witten,
  ``On The Structure Of The Topological Phase Of Two-Dimensional Gravity,''
  Nucl.\ Phys.\ B {\bf 340}, 281 (1990).}
\lref\dist{J.~Distler,
  ``2-D Quantum Gravity, Topological Field Theory And The Multicritical Matrix
  Models,''
  Nucl.\ Phys.\ B {\bf 342}, 523 (1990).}
\lref\vv{E.~Verlinde and H.~Verlinde,
  ``A Solution Of Two-Dimensional Topological Quantum Gravity,''
  Nucl.\ Phys.\ B {\bf 348}, 457 (1991).}
\lref\dijkwit{R.~Dijkgraaf and E.~Witten,
  ``Mean Field Theory, Topological Field Theory, And Multimatrix Models,''
  Nucl.\ Phys.\ B {\bf 342}, 486 (1990).}

\Title
{\vbox{\baselineskip12pt
\hbox{hep-th/0504229}}}
{\vbox{\centerline{From Free Fields to $AdS$ -- III}}}

\centerline{Rajesh Gopakumar\foot{gopakumr@mri.ernet.in}}

\centerline{\sl Harish-Chandra Research Institute, Chhatnag Rd.,}
\centerline{\sl Jhusi, Allahabad, India 211019.}
\medskip

\vskip 0.8cm

\centerline{\bf Abstract}
\medskip
\noindent

In previous work we have shown that large $N$ field theory amplitudes,
in Schwinger parametrised form, can be organised into integrals over the 
stringy moduli space ${\cal M}_{g,n}\times R_{+}^n$. Here we 
flesh this out into a concrete implementation of open-closed string duality. 
In particular, we propose that the closed 
string worldsheet is reconstructed from the unique Strebel 
quadratic differential that can be associated to (the dual of)
a field theory skeleton graph. We are led, in the process, to 
identify the inverse
Schwinger proper times ($\s_i={1\over \t_i}$) with the lengths
of edges of the critical graph of the Strebel differential. Kontsevich's
matrix model derivation of the intersection numbers in moduli space
provides a concrete 
example of this identification. It also exhibits how 
closed string correlators very naturally
emerge from the Schwinger parameter integrals. 
Finally, to illustrate the
utility of our approach to open-closed string duality, 
we outline a method
by which a worldsheet OPE can be directly extracted from the field theory 
expressions. Limits of the 
Strebel differential for the four punctured sphere play a key role.

\vskip 0.5cm
\Date{April 2005}
\listtoc
\writetoc

\newsec{Introduction}

The emergence of a closed string worldsheet from large $N$ gauge theory 
diagrams has been one of those insights that we have struggled 
to make precise in the last thirty years. 'tHooft's double line 
representation of feynman graphs \thft\ had made it pictorially plausible
that a two dimensional surface underlay these diagrams. More recently with
the AdS/CFT conjecture \malda \gkp \witads\ 
and other examples, it became clearer that 
open-closed string duality was the basic mechanism by which a closed
string theory emerged from the gauge theory. 
However, despite some improved understanding in specific examples 
\gv \ov \gr , we 
do not yet have a complete and concrete 
proposal of how the closed string worldsheet (and the CFT living on it)
arises for general field theory amplitudes. 

In earlier papers in this series \ffa \ffb , 
we have developed an approach to
open-closed string duality for general field 
theory amplitudes starting with the free field limit (See \ffstr\
for an overview. See also \akh\car\gor .). The 
intuition \ffa\ 
was that the Schwinger parametrised representation of field 
theory correlators is the right starting point to try and see the closed 
string worldsheet emerge from the field theory. This is because the 
Schwinger parameters are really the moduli of the worldline graphs
of the field theory (i.e. the open string picture)
and it should be these same moduli that should parametrise the 
shape of the emergent closed string worldsheet, on correctly implementing
open-closed string duality \foot{Another approach to seeing the worldsheet 
of the string in the light cone framework is that of Thorn and 
collaborators \thorn . See also \karch\ .}.

In \ffb\ this intuition was largely borne out. An 
appropriate reorganisation of free field theory $n$-point amplitudes
(of fixed genus $g$ in the 'tHooft sense)  
was made in terms of 
so-called skeleton graphs. The integral over the 
effective Schwinger parameters
together with the sum over inequivalent skeleton graphs was argued to
be actually an integral over the {\it stringy} moduli space
${\cal M}_{g,n}\times R_{+}^n$. Here ${\cal M}_{g,n}$ is the moduli 
space of genus $g$ surfaces with $n$ {\it punctures} -- exactly what one
would expect for the closed string dual. 
This identification crucially used the fact that the space of 
these skeleton graphs (or rather their dual), together with a positive
length assignment (the Schwinger time) for each edge, gives a   
natural cell decomposition of ${\cal M}_{g,n}\times R_{+}^n$. This cell
decomposition, which 
is originally due to Mumford, Harer, Penner,
Kontsevich \harer\pen\kont\ and other mathematicians,
is also natural from the point of view of
cubic open string field theory \gmw\zwie . That the
space of Schwinger parameters of skeleton graphs is isomorphic to a 
stringy moduli space is a very encouraging sign 
that we are implementing open-closed string duality. It is 
evidence that a closed string theory {\it can} very naturally, and rather 
generally, emerge from the gauge theory amplitudes. 

We would, of course, like to do more. The identification of the 
space of  Schwinger parameters and skeleton graphs with the stringy 
moduli space implies that the field theory expression for 
the {\it integrand} over the Schwinger parameter space is to be identified 
with a correlator of closed string vertex operators. We would 
therefore like to 
extract from these expressions the properties of
the putative closed string worldsheet CFT. For this we would need 
a dictionary between the Schwinger parameters of the 
field theory skeleton graphs and the usual complex coordinates
on ${\cal M}_{g,n}$ in terms of which, for instance, the holomorphic 
properties of CFT correlators is made manifest. Strictly speaking, 
we would use
such a dictionary to first {\it check} whether the integrand in the Schwinger 
parameter space indeed satisfies all the requirements
for a worldsheet CFT correlator. In any case, arriving at 
such a dictionary entails making 
a definite identification of the Schwinger parameters with parameters 
specifying the closed string worldsheet. It is therefore the ingredient 
that would  
make our proposal 
for implementing the {\it geometry} of open-closed string duality complete.   

One of the goals of
the present work, is to arrive at this precise dictionary. 
We will, in fact, propose a concrete method to reconstruct the 
particular closed string worldsheet corresponding to a given point in the 
Schwinger parameter space of skeleton graphs. 
The identification makes use of the mathematical 
correspondence which underlies the cell decomposition of 
${\cal M}_{g,n}\times R_{+}^n$ that was mentioned above. This correspondence
proceeds via the construction of certain unique quadratic differentials,
known as Strebel differentials, on a Riemann surface. The so called 
critical graphs of these Strebel differentials will be identified with the 
dual of the field theory skeleton graphs. The crucial ingredient will be 
the dictionary between the Schwinger proper times ($\t_r$)
and the lengths ($l_r$) of the 
edges of the critical graph. The $l_r$ are important since
they give a unique parametrisation 
of the closed string worldsheet. In particular, they are 
determined in terms of the complex coordinates on ${\cal M}_{g,n}$
denoted collectively by $z_a$, together with additional data of the
$ R_{+}^n$.
We will argue 
that the relation between the Schwinger times and the Strebel lengths 
is simply 
\eqn\dic{\s_r\equiv {1\over \t_r}=l_r(z_a).}
We will see that this identification is natural from many points
of view. The general picture of open-closed string duality that then 
emerges is very much in line with one's intuition of field theory Wick 
contraction lines being glued up to form the closed string worldsheet. 
It is also in line with various bit pictures of the closed string 
worldsheet. Interactions in the field theory 
are also readily incorporated into this picture since they 
correspond to insertions of additional closed string vertex operators.

We will also revisit Kontsevich's classic 
derivation \kont\ of Witten's conjecture \witint\ on the intersection numbers 
in moduli space. It will illustrate for us, in a concrete way, 
how the Schwinger parametrisation provides the natural passage 
to the dual closed string. Moreover, it will also  
exhibit how integrands in the Schwinger 
parameter space become correlators on ${\cal M}_{g,n}$.

All this adds up to a satisfying picture of the way the closed string 
worldsheet emerges from the field theory. But as mentioned above, one of the 
aims of getting a precise dictionary
is to read off expressions for worldsheet CFT correlators in terms of the 
usual complex parameters $z_a$ (such as for the locations of punctures).
Thus we would like to re-express the 
field theory integrand, which can be written in terms of $\s_r$, 
in terms of the $z_a$ using \dic . In general, \dic\ implies a complicated 
transcendental relation between the $\s_r$ and the $z_a$. This 
is because, as we shall review, the relation between the Strebel lengths
$l_r$ and the $z_a$, while precisely defined, can be analytically involved
in general.

We will therefore take the strategy of looking for simplifications 
at the boundary of moduli space. The instance of the four 
point function on the sphere 
is the simplest non-trivial case to consider. We will 
focus on the limiting Strebel differentials that arise when two 
punctures come together on ${\cal M}_{0,4}$. 
From the field theory point of view, 
we can zoom in on this region by considering 
particular UV limits of the free field spacetime four point correlator. 
Basically, the idea is that as we take two points in spacetime 
close to each other, the field theory integrand gets all its
contribution from the corresponding proper time interval $\t\R 0$. This
translates into limiting behaviour that we expect for the Strebel 
differential near boundaries of ${\cal M}_{0,4}\times R_+^4$.

In other words, {\it the worldsheet operator product expansion
originates from the spacetime
operator product expansion}. Indeed, we can systematically
write the short distance expansion of spacetime correlators 
in a Schwinger parametrisation. As mentioned, this will be an expansion 
in the proper times $\t$ that go to zero in this limit. The dictionary  
\dic\ together with the characterisation of the limiting Strebel
differential in this region of ${\cal M}_{0,4}$ gives a well-defined
method to convert the latter expansion into a regular worldsheet expansion 
in terms of a usual complex coordinate $z\R 0$. This provides us the precise 
setting to check whether the Schwinger integrand satisfies all the
requirements of a worldsheet CFT correlator. 
Though we will reserve a  
detailed study of this issue for later, here 
we will use the scaling behaviour of the limiting Strebel differential 
to deduce that the worldsheet expansion is actually in powers of  
$|z|^{1 \over 2}$. This is very encouraging as it might be a signature of 
an underlying {\it fermionic} string.

An outline of the organisation of the paper: In the next section 
we briefly review the reorganisation \ffb\ of Schwinger parametrised 
field theory amplitudes into skeleton graphs. In Sec. 3,
we review the notion of a Strebel 
differential, its critical graph and Strebel lengths, 
as well as some of the important results regarding 
these objects. What will play an important role for us is how 
the Strebel differentials give rise to a natural cell decomposition of 
${\cal M}_{g,n}\times R_{+}^n$. In Sec. 4, we detail the connection
of this cell decomposition 
to the space of field theory skeleton graphs with its Schwinger parameters.
We give a couple of arguments for the dictionary \dic . We also describe why
the resulting picture implements open-closed string duality as expected.
We revisit Kontsevich's derivation in Sec.5. 
We also see here
how closed string correlators naturally arise from the Schwinger 
integrand. In Sec. 6 we take the first steps to using our 
open-closed dictionary. We first explain the mathematical steps 
in going from the Strebel parametrisation of moduli space to the usual 
holomorphic one. Applying this to the four point function, we see how 
to go to the boundary of moduli space by 
taking various UV limits of the spacetime correlator. We study the Strebel
differential in these limits and show by scaling arguments that the 
resultant worldsheet expansion hints at an underlying fermionic string 
theory. In Appendix A we collect some useful explicit 
expressions for the Schwinger parametrisation of certain four point 
functions.

\newsec{Skeleton Graphs and Field Theory Amplitudes}

The idea behind reorganising field theory diagrams into 
skeleton graphs \ffb\ is quite simple. Any perturbative gauge invariant
correlator 
can be written in terms of double lines (assuming only fields 
in the adjoint representation) with some number of vertices as well as 
wick contractions between the fields at each vertex.  These vertices 
could be either internal or external. We will assume that the graphs 
have already been organised according to their 't Hooft genus. 
The large $N$ limit gives a way to separate out the contributions 
of different genera via a natural small expansion parameter \thft .  

At the graphical level, the skeleton
graph associated to any such gauge theory diagram 
is simply the graph obtained by merging together all the homotopically
equivalent contractions between any two pairs of vertices. 
By homotopically equivalent, we mean those double line contractions that can 
be deformed into each other without crossing any other line or vertex. 
The nett result is that we will not have any faces which are bounded by 
only two edges (contractions). Each face of the skeleton graph will have 
at least three edges. In fact, the generic situation for correlators of 
composite operators with enough elementary fields will be to have
all faces triangular. Thus the generic skeleton graph will be a triangulation 
of the genus $g$ surface with as many vertices as there are internal and 
external ones. 

\fig{Gluing up of a planar six point function into a skeleton graph.}
{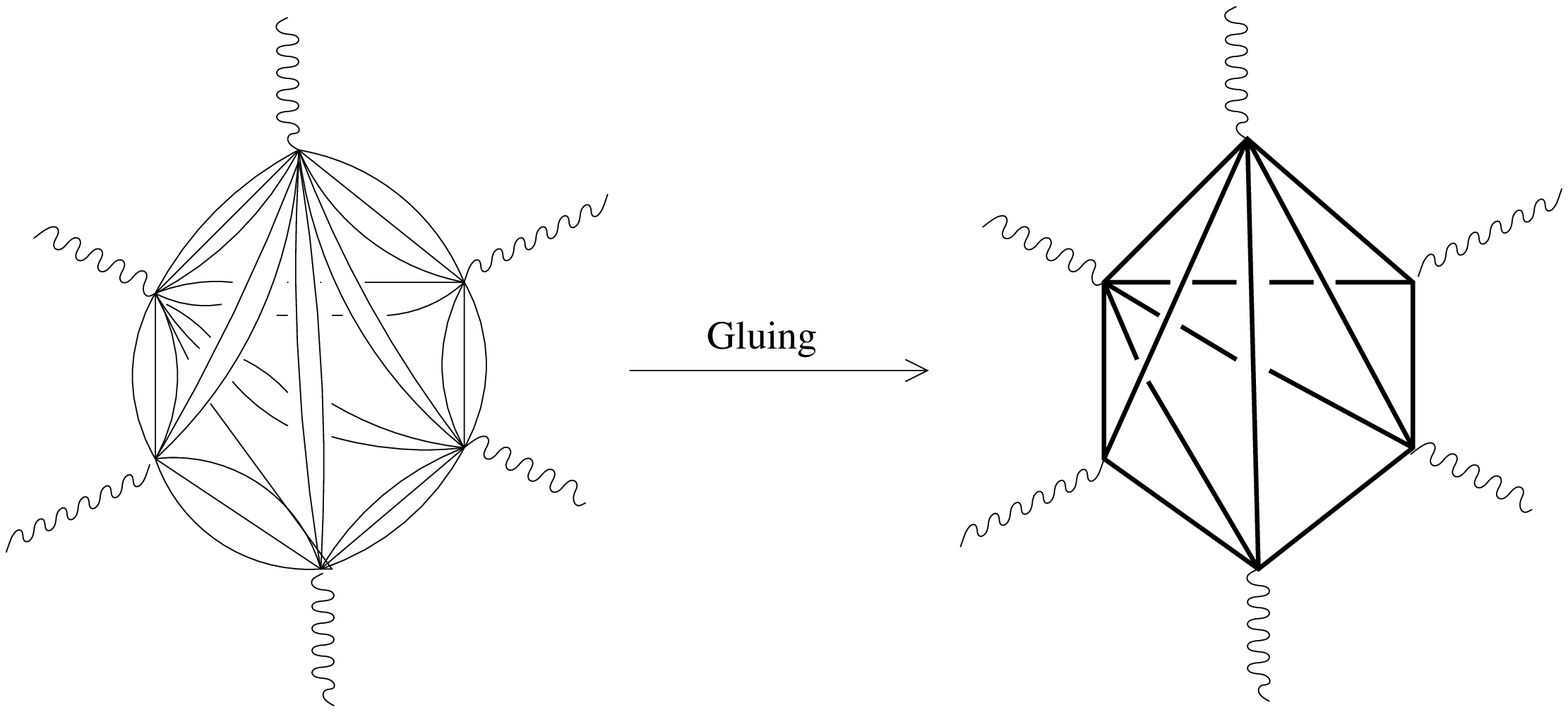}{6.0truein}

This partial gluing up of the field theory contractions is something one
would expect from open-closed string duality. It is the first stage of 
closing up all the holes of the open string worldsheet. In \ffb\
we saw that this gluing is reflected very nicely
in the Schwinger parametrised expressions for correlators. Let us briefly
review the logic. For definiteness,
and simplicity, we considered a free field 
$n$-point function built out of adjoint scalars 
\eqn\ncorr{
G^{\{J_i\}}_{g}(k_1, k_2,\ldots k_n)=
\la \prod_{i=1}^n \Tr\Phi^{J_i}(k_i)\ra_{g}. }
The subscript indicates that we are considering the contributions
of genus $g$.  

In terms of Schwinger times $\tt$ for each propagator, and after carrying out 
the integral over the $d$-dimensional 
internal momenta, one obtained an expression of the form 
\eqn\ttpar{G^{\{J_i\}}_g(k_1, k_2,\ldots k_n)
=\sum_{graphs}\int_0^{\infty}{[d\tt]\over \D(\tt)^{d\over 2}}
\exp[-P(\tt,k)].}
There are definite graph theoretic 
expressions for $P(\tt,k)$ and $\D(\tt)$ whose details  
need not concern us at present. The main point is that though this
is an expression depending on as many Schwinger times $\tt$ as there are 
internal propagators to the graph, there is actually a vast simplification. 
The above expression, for any given graph, can be written purely
in terms of the 
skeleton graph associated to that graph, where one assigns the
effective Schwinger time 
\eqn\resist{{1 \over \t_r} =\sum_{\m_r=1}^{m_r}{1\over \tt_{r\m_r}}.}
Here $r$ labels an edge of the skeleton graph obtained by merging 
the homotopic set of $m_r$ edges, indexed by $\m_r$. 
After making the change of variables \resist\ to the 
contributions in \ttpar\ we get
\eqn\skel{\int_0^{\infty}{\prod_{r,\m_r} d\tt_{r\m_r} \over \D(\tt)^{d\over 2}}
e^{-P(\tt,k)} =C^{\{m_r\}}\int_0^{\infty}
\prod_r({d\t_r\over \t_r^{(m_r-1)({d\over2}-1)}}){e^{-P_{skel}(\t, k)}
\over \D_{skel}(\t)^{d\over 2}}.}
$P_{skel}(\t, k)$ and $\D_{skel}(\t)$ are given in terms of the graph
connectivity of the skeleton graph with the assignment $\{\t_r\}$ to 
its edges. The $C^{\{m_r\}}$ is a numerical factor
coming from the Jacobian of the change of  variables and is 
computed in \ffb . 

As a result the entire expression for the $n$-point function can be 
expressed completely
as a sum over all the skeleton graphs that contribute to the 
amplitude.
\eqn\schem{G^{\{J_i\}}_g(k_1, k_2,\ldots k_n)=\sum_{skel.\atop graphs}
\int_0^{\infty}
{\prod_r d\t_r f^{\{J_i\}}(\t)\over\D_{skel}(\t)^{d\over 2}}
e^{-P_{skel}(\t, k)}.}
The $f^{\{J_i\}}(\t)$ come from carrying out the sum over the
multipicities $m_r$ that are compatible with the same skeleton graph and 
the nett number of fields ${\{J_i\}}$. All the $J_i$ dependence resides 
in this term. 
Its explicit form is again available
in \ffb . 
The sum in Eq.\schem\ 
is then over the various inequivalent (i.e. with inequivalent connectivity)
skeleton graphs of genus $g$ with $n$ vertices. 

This partial gluing up has accomplished a big simplification of the 
Schwinger parametrised representation. In particular, an
important point to note is that the number of edges (and thus effective
Schwinger times $\t$) in the skeleton
graph only depends on $n$ 
and $g$ (and {\it not} the $J_i$). In fact, for a generic triangulation, 
the number of edges is $6g-6+3n$. 
The universality of this representation, 
so to say, is the reason why the moduli space of skeleton graphs is the more
natural object to consider. It is the space which will provide the cell
decomposition of ${\cal M}_{g,n}\times R_+^n$, to which we now turn.

\newsec{Strebel differentials and the Cell Decomposition of  
${\cal M}_{g,n}\times R_+^n$}

To make the connection between the field theory skeleton graphs and 
${\cal M}_{g,n}\times R_+^n$, we will need to take a small excursion
into the topic of quadratic differentials \streb\ on Riemann surfaces. 
For a very nice and clear exposition of much of the material in this 
section see \mp . See also \sunil\ for a physicist's review of some 
general facts about quadratic and Strebel differentials.

\subsec{Quadratic Differentials}

On any Riemann surface, we will be interested in
considering meromorphic quadratic differentials. These take the
form $\phi(z)dz^2$ in any complex coordinate chart (parametrised by $z$),
with $\phi(z)$ being a meromorphic function of $z$ in that chart.
Under a holomorphic change of coordinates to $w=w(z)$ we have
\eqn\qdtr{\phi(z)dz^2=\tilde{\phi}(w)dw^2 \Rightarrow 
\tilde{\phi}(w)=\phi(z(w))({dz\over dw})^2.}

Something that will be important for us is 
that given a meromorphic quadratic differential, we can use it to 
define a (locally) flat metric on the Riemann surface.
This is simply given by the line element
\eqn\qdmet{ds^2=|\phi(z)|dzd\bar{z}.}
This is well defined away from the zeros and poles of $\phi(z)$.
 
Another crucial notion is that 
of horizontal and vertical trajectories. Consider a 
curve $z(t)$ in a coordinate chart, where $t$ is a parameter along the 
curve taking values in some interval of the real line. A curve $z_H(t)$ is 
called horizontal with respect to a given quadratic differential $\phi(z)$
if 
\eqn\hor{\phi(z_H(t))({dz_H\over dt})^2 >0,}
for all $t$ in the interval. 
Similarly, a curve $z_V(t)$ is vertical if 
\eqn\ver{\phi(z_V(t))({dz_V\over dt})^2 <0,}
again for all $t$ in the relevant interval. 

The terminology is motivated by the simplest such differential on $C$, namely
$dz^2$. Then it is obvious that the 
horizontal trajectories are simply all horizontal lines (i.e. parallel to 
the real axis). and the vertical trajectories are all vertical lines, 
parallel to the imaginary axis. In fact, at a {\it regular} point 
on the Riemann
surface, by a suitable change of coordinates, we can always put a general 
quadratic differential to be of the form $dz^2$ in the infinitesimal 
vicinity of that point. Then the horizontal and vertical trajectories in 
the vicinity of the point form a rectangular grid like in $C$. 

Near a zero or a pole, however, the behaviour is quite different. 
Any quadratic differential,
in the vicinity of a zero or pole can be put in the form 
\eqn\zp{\phi(z)dz^2=z^mdz^2.}
Here $m$ is an integer, positive for a zero and negative 
for a pole. However, a double pole, i.e. $m=-2$, will have to be
treated specially. 
It is easy to verify (for $m\neq -2$) that the radial half lines
\eqn\radhor{z_H(t)=t\exp{({2\pi ik\over m+2})}; ~~~~~~ t>0,
~~~~~(k=0\ldots m+1)}
are horizontal trajectories. 
And that
\eqn\radver{z_V(t)=t\exp{({\pi i(2k+1)\over m+2})}; ~~~~~~ t>0,
~~~~~(k=0\ldots m+1)}
are vertical trajectories.
Note that near a simple zero ($m=1$), we have three horizontal (as well
as three vertical trajectories) meeting at the location of the zero. 
A double zero has four horizontal trajectories intersecting etc.   

The case of a double pole, $m=-2$, is special. Consider a quadratic 
differential 
in the neighbourhood of such a pole and taking the form 
\eqn\dpole{\phi(z)dz^2=-{p^2\over (2\pi)^2}{dz^2\over z^2}.}
Notice that by a single valued 
change of coordinates we cannot change the 
coefficent multiplying the double pole. We will denote this invariant 
coefficent (in this case $-{p^2\over (2\pi)^2}{dz^2\over z^2}$)
as the residue of the quadratic differential at the double pole. 
The reason for the above parametrisation of the residue will become 
clear very soon. 

We will in fact only need to consider $p$ to be real and positive. 
In this case,
it is easy to work out 
that the horizontal trajectories are concentric circles about the pole
i.e. 
\eqn\dphor{z_H(t)=re^{it}; ~~~~~~~ t\in (0,2\pi)}
with $r$ an arbitrary constant. Meanwhile, {\it any} radial line
emanating from $z=0$ 
\eqn\dpver{z_H(t)=te^{i\theta}; ~~~~~~ t>0,}
with $\theta$ fixed 
is a vertical trajectory.   

With respect to the flat metric defined by \qdmet\ we see that 
near a double pole, the circular horizontal trajectories in \dphor\ 
all have equal circumference $p$. This explains the parametrisation 
of the residue in \dpole . Notice that the pole itself is at an 
infinite distance from any finite point.
The geometry of the Riemann surface 
near a double pole is thus that of a semi-infinite cylinder of 
circumference $p$ in which the 
horizontal trajectories are the circular cross-sections while the 
vertical trajectories are parallel to the axis of the cylinder.

We will be interested, in what follows, in quadratic differentials
with only double poles. 

\subsec{Strebel Differentials}

So far we have discussed the {\it local} structure of horizontal and 
vertical trajectories of a general meromorphic quadratic 
differential. The {\it global} structure of these trajectories is actually
very interesting. 

Consider a Riemann surface $\Sigma_{g,n}$ of genus $g$ with $n$ 
marked points (which we will identify with the punctures).
Generically, a horizontal trajectory of a quadratic differential on
$\Sigma_{g,n}$ will wander around the surface without closing 
on itself. However, there 
are {\it special} quadratic differentials on $\Sigma_{g,n}$
for which essentially all horizontal trajectories are closed curves. 
These special differentials have the following properties:
They have {\it only}  double poles which are all at 
the locations of the marked points. The residues at these double poles, 
as defined in \dpole ,  have 
$p$ real and positive. The set of closed 
horizontal trajectories around each such pole foliate a punctured disc 
about this pole (a so called maximal ring domain). The boundary of the disc 
contains a certain number of zeros 
of the quadratic differential. In fact, the boundary  
is a union of the {\it non-closed} 
horizontal trajectories whose endpoints are the zeroes. Moreover, 
the entire Riemann surface is a union of the $n$ ring domains about each 
marked point (double pole) together with the non-closed horizontal 
trajectories which comprise the boundary of these discs.  
Because of a theorem of Strebel stated below, we will call such a 
special quadratic differential, a Strebel differential. 
A ring domain of a Strebel differential near a double pole is shown 
in Fig.2. 

The result of Strebel \streb\ states that for every Riemann surface 
$\Sigma_{g,n}$ (with $n>0$ and $2g+n>2$) {\it and} any 
$n$ specified positive numbers
$(p_1,p_2,\ldots p_n)$, there exists a {\it unique} 
Strebel differential. Namely, a quadratic differential that 
is holomorphic everywhere on $\Sigma_{g,n}$ except for the 
$n$ marked points where it has double poles with residue determined by
$p_i$ at the $i$'th pole. It has the property that the ring domains about 
the poles, foliated by the closed horizontal trajectories, cover 
the entire surface. The measure zero set of boundaries 
of the ring domains, comprise of the non-closed 
horizontal trajectories 
that begin and end on zeros \foot{This theorem of Strebel was used
by \saadi\kugo\ to construct the higher order contact vertices in 
{\it closed} string field theory. In that context the residues 
were fixed to be all equal to $2\pi$. Moreover, since the contact vertices 
were supposed to only cover the part of moduli space missed by the 
Feynman diagrams from lower order vertices, certain inequalities were 
also imposed on the lengths of closed cycles. This description of 
closed string field theory has been used to study quartic interactions 
\belzw\bel\moel\yangz .}.

\fig{A characteristic ring domain in the vicinity of a double pole (marked
with a dot). The non-closed horizontal trajectories are shown by thick lines.
These begin and end at zeros marked by a cross.}
{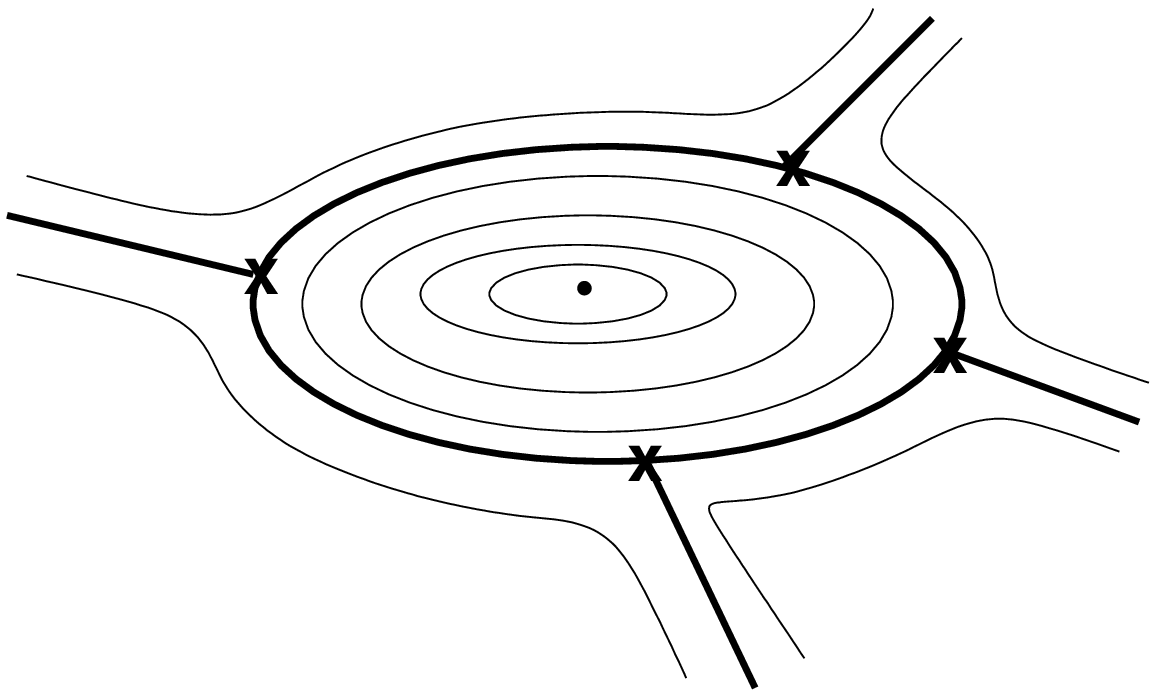}{3.0truein}

\subsec{The Cell Decomposition of ${\cal M}_{g,n}\times R_{+}^n$}

What we have just stated is that
associated with every point on the extended (or ``decorated'') moduli space
${\cal M}_{g,n}\times R_{+}^n$, is a unique Strebel differential. 
This result gives rise to a nice cell decomposition of 
${\cal M}_{g,n}\times R_{+}^n$ which will then connect up with our
field theory discussion. 

First, notice that the set of non-closed horizontal trajectories of a 
Strebel differential (i.e. the boundaries of the ring domains), on 
$\Sigma_{g,n}$,
forms a graph which is embedded into the Riemann surface. 
By imagining the edges of the graph to be thickened (a double line or 
``ribbon'' graph)
we see that we can, as usual, associate a genus $g$ to this graph. 
This is called the {\it critical graph} of the corresponding Strebel
differential.   
Since,
as we mentioned, the edges of this graph connect various zeros 
of the Strebel differential, the vertices where they meet
are generically trivalent (corresponding to simple zeros). Because the 
ring domains are all punctured discs and are $n$ in number, the resulting 
graph has $n$ (polygonal) faces. A simple application of $V-E+F=2-2g$ for the 
generic trivalent graph implies that the number of edges of the 
critical graph is $E=6g-6+3n$. The number of vertices, which is also
the number of (simple) zeros is $V=4g-4+2n$. 

We can also assign a length to each edge of the critical graph. For an edge
connecting zeros $z_i$ and $z_j$, the length is given by 
\eqn\strl{l_{ij}=\int_{z_i}^{z_j}\sqrt{\phi(z)}dz}
where the contour is chosen so that it is homotopic to the non-closed 
horizontal trajectory connecting $z_i$ with $z_j$. 
Note that the value of the integral is then 
independent of the particular contour and is actually real positive 
since it is real positive along the horizontal trajectory connecting the
two zeros. Given these length assignments we can define a combinatorial
space ${\cal M}_{g,n}^{comb}$, following Kontsevich. It is the space of all
ribbon graphs of genus $g$ with $n$ marked faces, with all vertices at least 
trivalent and a length assigned to each edge. Each inequivalent 
ribbon graph with trivalent vertices fills out a top dimensional cell
in ${\cal M}_{g,n}^{comb}$. These different cells  
connect with each other when one edge collapses to zero length in the 
$s$-channel, so to say. One can go to the adjacent cell
by expanding out the collapsed vertex but
now in the $t$-channel.  

It turns out that ${\cal M}_{g,n}^{comb}$ gives a cell
decomposition of ${\cal M}_{g,n}\times R_+^n$.
Let us describe the isomorphism between ${\cal M}_{g,n}^{comb}$
and the extended moduli space ${\cal M}_{g,n}\times R_+^n$.
Given a point in ${\cal M}_{g,n}\times R_+^n$, we can immediately 
associate a point in ${\cal M}_{g,n}^{comb}$ via the result of Strebel.
We just construct the unique Strebel differential on the corresponding 
surface $\Sigma_{g,n}$ (with specified residues at the $n$ poles) and 
find its critical graph. This is a ribbon graph of genus $g$ with $n$ 
marked faces and we can assign to the edges of this graph the Strebel 
lengths defined in \strl . This gives us a unique point in 
${\cal M}_{g,n}^{comb}$.

The reverse map is more interesting for us. Given a point 
in ${\cal M}_{g,n}^{comb}$ (i.e. a ribbon graph 
with specified lengths of the edges), there is a canonical way to construct a 
Riemann surface $\Sigma_{g,n}$. 
The geometric picture is that $\Sigma_{g,n}$ consists 
of semi-infinite flat cylinders glued  
onto each face of the given graph (in an oriented way). The circumference
of the cylinder on the $i$'th face is given by 
the sum of all the lengths of the edges bordering that face.  

There is actually a systematic way of constructing $\Sigma_{g,n}$ from
the ribbon graph, which will be  
important for open-closed 
string duality. 
We will not go into the details of the construction here,
which can be found, for instance, in \mp , but rather sketch the main 
elements involved.
For each edge of the ribbon graph of assigned 
length $l_{ij}$, we construct 
an infinite strip of uniform 
width $l_{ij}$ aligned parallel to the imaginary axis in $C$. 
On this strip we 
have the differential $dz^2$ in terms of the natural coordinate.
How do we glue all these strips into a surface? 
When three or more edges meet at a vertex, we patch the corresponding 
strips together  
in a way familiar from open string field theory. For instance, for a 
trivalent vertex, in some neighbourhood of the vertex we use a new 
coordinate $w\propto z^{2\over 3}$. This maps the part of each strip in the
vicinity of the 
vertex into a wedge of angle ${2\pi \over 3}$. The three wedges 
are glued together in the $w$-plane (see Fig.3). 
The individual differentials $dz^2$
for each strip are now transformed into a differential of the form 
$wdw^2$ in the $w$-plane which smoothly overlaps between 
the different strips in the vicinity of a vertex. 
For a general $k$-valent vertex of the ribbon graph, 
we use a $z^{2\over k}$ map for gluing
the strips together.

\fig{Three strips glued in the vicinity of a trivalent vertex. The thick lines
are the {\it horizontal} 
trajectories in the middle of each strip. The thin lines
are representative {\it vertical} trajectories.}
{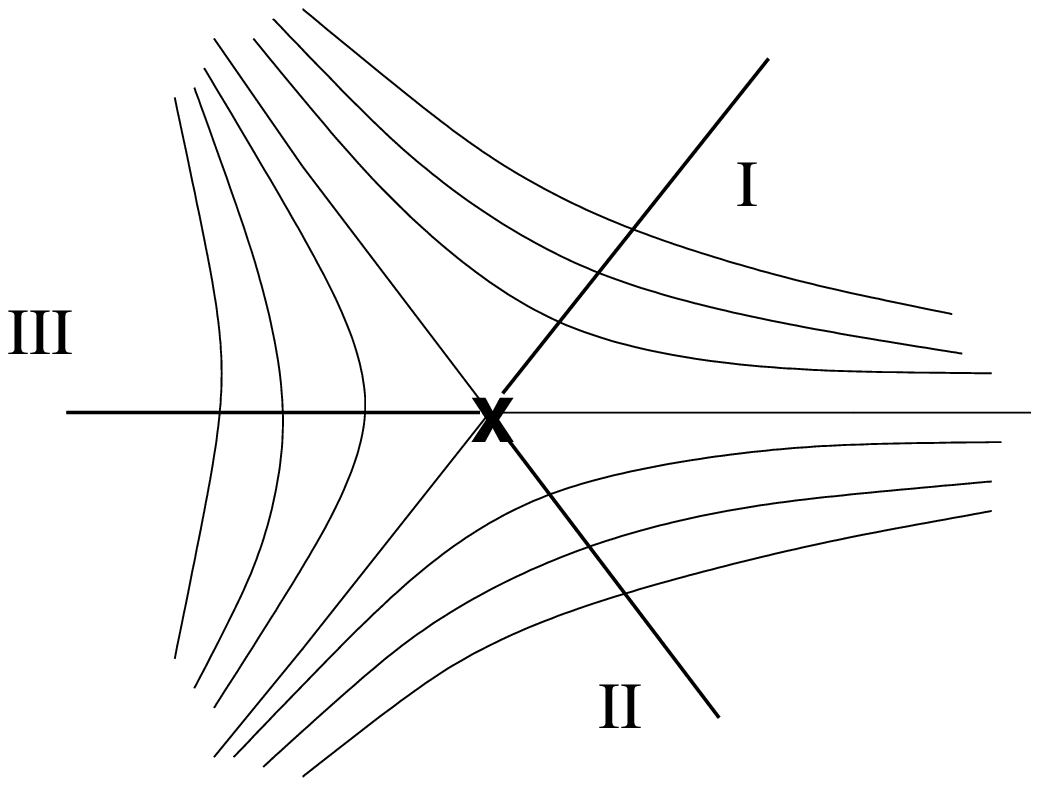}{2.5truein}

Similarly, when we have several edges bounding a face, we glue 
halves of the corresponding strips so that they form a cylinder (see Fig.4).  
This can be done  
by an exponential map $u(z)\propto \exp{{2\pi iz\over p_i}}$
which transforms the differentials $dz^2$ in the individual strips 
into a differential of the form 
$$\tilde{\phi}(u)du^2=-{p_i^2\over (2\pi)^2}{du^2\over u^2}.$$
Here $p_i$ is the sum of the lengths of all the
edges bounding the $i$'th face and 
the map is such that as the argument of $u$ goes over $(0,2\pi)$, one goes
over all the edges in the corresponding order. For explicit details of these
maps see \mp . 

Having specified the complex coordinates in each strip and the 
gluing rules at vertices and the centre of faces, one can show that
one has constructed a  unique
Riemann surface $\Sigma_{g,n}$. Moreover, the quadratic 
differential obtained by thus
gluing together the ones in each chart is the unique Strebel differential
for this surface with residues given by $\{p_i\}$. This Strebel 
differential therefore captures all the information about the point in
${\cal M}_{g,n}\times R_+^n $ that we obtain from this construction. 
Combined with the previous map from ${\cal M}_{g,n}\times R_+^n$
to ${\cal M}_{g,n}^{comb}$, we have described 
an isomorphism between the two spaces. 

\fig{Half-strips glued together into cylinders. Only one full strip of width
$l$ is shown. Figure adapted from \zvon .}
{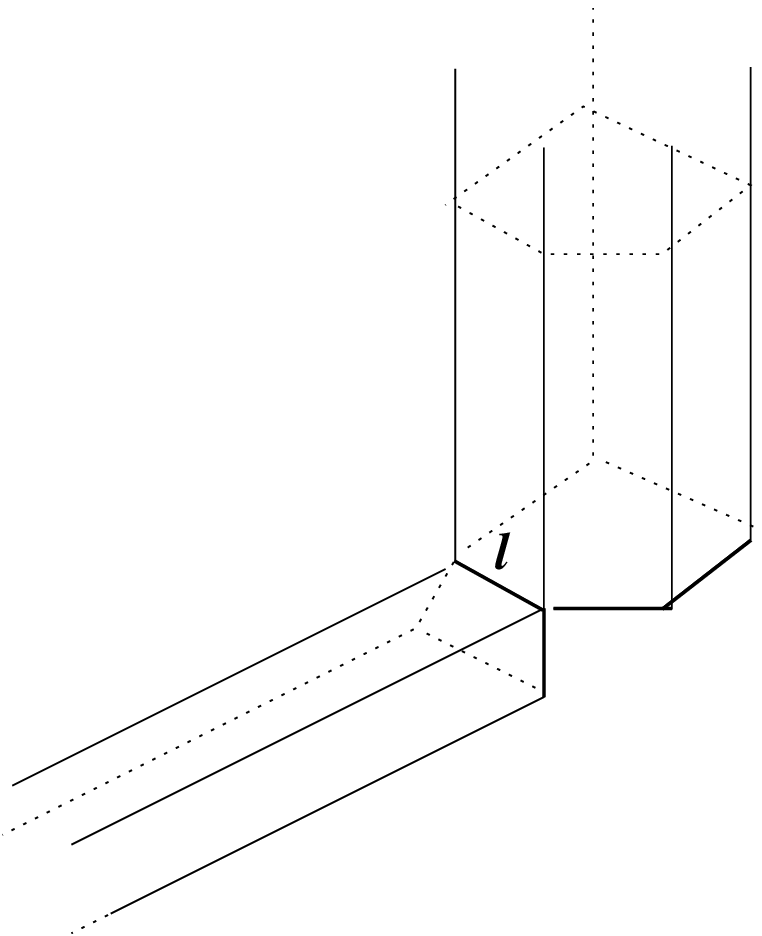}{2.5truein}

What is of primary interest to us is that we have  
a definite mathematical procedure to build up
a closed string surface of genus $g$ with $n$ punctures entirely 
out of flat strips glued together, starting from the data in the 
ribbon graph. In the next section, we will argue that this construction
is exactly how open-closed string duality is implemented. 

\newsec{Schwinger Times and Strebel Lengths}

In \ffb\ it was observed that the space of inequivalent skeleton graphs
(with a Schwinger time associated to each edge) is isomorphic to the space 
${\cal M}_{g,n}^{comb}$ described above. This is because one can consider 
the space of dual graphs to the skeleton graphs considered in Sec.2. 
The dual graphs also have genus $g$ but $n$ faces. The generically triangular  
faces of the skeleton graph go over to generically trivalent vertices 
in the dual graph. These are exactly the set of ribbon graphs that appear 
in ${\cal M}_{g,n}^{comb}$. Since the original edges were assigned a 
parameter $\t_r$, and since there is a dual edge for each edge, 
we also have a length assignment for each edge of the dual graph.
Thus we have an isomorphism 
from the moduli space of skeleton graphs to ${\cal M}_{g,n}^{comb}$.

Using the further isomorphism between ${\cal M}_{g,n}^{comb}$ and 
${\cal M}_{g,n}\times R_+^n$ described in the previous section, 
it was 
argued in \ffb\ that the field theory expression \schem\ was actually 
therefore an
integral over ${\cal M}_{g,n}\times R_+^n$. While this argument is adequate 
to understand how field theory diagrams can 
reorganise into string amplitudes, it is not sufficent for 
extracting useful information about the dual string theory.

For that one has to have a precise dictionary between the 
Schwinger times and the string moduli. 
Since the isomorphism between the Schwinger parameter space 
and the string moduli space
was established via ${\cal M}_{g,n}^{comb}$, we basically need
a relation between  
the Schwinger times and the Strebel lengths 
which parametrise ${\cal M}_{g,n}^{comb}$. 

We will now argue that this is simply given by \dic\ i.e. we identify 
the inverse Schwinger times $\s_r$ with the Strebel lengths $l_r$ 
\foot{Note that, mathematically, there is no real restriction
on the functional relation between $l$ and $\s$.
Any monotonic function $f$ such $f(0)=0$ and $f(\infty )=\infty$ 
would fit the bill as far as the isomorphism between the moduli 
space of skeleton graphs and ${\cal M}_{g,n}^{comb}$ is concerned.}. 
A heuristic argument for this identification is as follows.
View the field theory worldline as an open string worldsheet 
whose width is going to zero. In fact, let us regularise the width to be 
$\ep$ with the understanding that it is to be eventually taken 
to zero. The length of this line is the Schwinger time $\t$. 
By means of a uniform conformal transformation on the worldsheet, we 
map this to a flat worldsheet of width $\s={1\over \t}$ and length
${1\over \ep}$ (see fig. 3). 
In the limit as $\ep \R 0$, this is an infinite 
strip of fixed width $\sigma$. 
{\it We will identify these flat infinite strips
with the ones that appeared in the gluing construction of the
previous section}. These formed the building blocks for the 
closed Riemann surface. There we saw that the widths
of the strips were 
the Strebel lengths of the closed string Riemann surface.
By our identification of the open string worldsheet with these strips
we see that $l=\s={1\over \t}$. 

\fig{A worldline of length $\t$ is equivalent to an infinite strip of
width $\s={1\over \t}$.}
{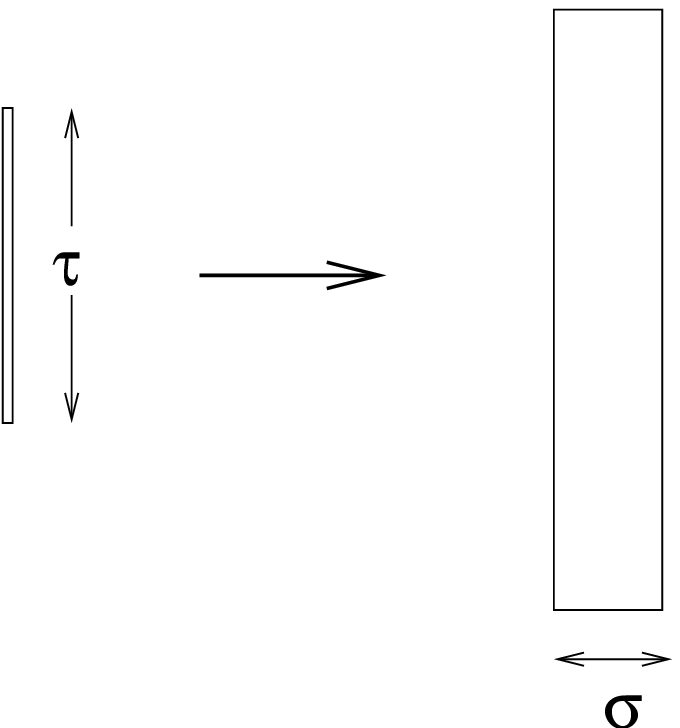}{2.0truein}

Another, somewhat complementary, 
way to arrive at this identification of $\s$ with the 
Strebel lengths is to notice that in the field theory, when we 
partially glued together homotopically equivalent lines, the effective 
Schwinger parameters were given by \resist . In terms of 
the ``conductances'' this simply means that the $\sigma$'s are 
{\it additive} on gluing. Now notice 
that the Strebel lengths are also additive when  
we glue two parallel strips with each other. In constructing the 
Riemann surface, we simply place two such strips parallel to each other. 

Therefore, since the 
functional relation between the $\s$ and the $l$ respects this additive 
property, the two have to be a multiple of each other. We can always 
set that multiple to one by a change of scale. This again suggests 
the dictionary \dic . In addition to both these arguments, we will also 
see in the next section
that this identification is consistent with Kontsevich's  derivation.

Note that having made the physical identification 
between the Schwinger times with the Strebel lengths, 
the field theory integrand, such as in \schem , is now expressible
directly in terms of parameters of the closed string. The parametrisation 
in terms of Strebel lengths is perhaps an unfamiliar parametrisation of 
${\cal M}_{g,n}$, though it suggests a natural string field 
theory origin. In any case, as mentioned
in the introduction, to put the integrand into familiar form, we should
construct the map  
between the $l_r$ and the usual complex moduli $z_a$
of ${\cal M}_{g,n}$. 
This is a well defined map since it is 
constructed via the Strebel differential.
What is to be stressed is that the construction of this map
is a purely mathematical
question. We will return to it in Sec. 6.

The arguments of this section
add up to a coherent picture of how open-closed string
duality is being implemented. 
The original double line wick contractions 
in the field theory, we have seen, are equivalent 
to infinite open string strips with width $\tilde{\s}={1\over \tilde{\t}}$. 
Homotopically
equivalent wick contractions are easily glued onto each other and the 
width ($\s_r$) of the resultant strip
is just the sum of the individual widths ($\tilde{\s}_{r\m_r}$). 
This is the stage where 
we obtain the skeleton graph with effective Schwinger parameters. The 
strips corresponding to the skeleton graph are now glued together as
described in the previous section. We now also see why 
the dual to the skeleton graph played a role in \ffb .
The open string strips are parallel to the
{\it vertical} trajectories of the Strebel differential of the 
closed string worldsheet. It is therefore the  
dual graph to the skeleton graph which is in correspondence with the 
(non-closed) {\it horizontal} trajectories which form the critical graph 
of the Strebel differential. 

That strips of varying sizes are getting glued up also goes well 
with a bit picture in which bit worldlines combine to form
the closed string worldsheet. It would be nice to make 
this more precise, perhaps in a lightcone gauge or as in the 
recent work of Alday et.al. \justin .

Though we had mainly discussed correlators in the 
context of free fields, it is clear 
that the process of open-closed duality we have described here is more
general. We can consider an arbitrary field theory diagram in 
perturbation theory and carry out the procedure described in Sec.2 
and over here. All that changes is that the internal vertices 
give rise to additional marked points or punctures 
on the Riemann surface. These punctures would have closed 
string vertex operator insertions just as with the external punctures.   
We will need to exponentiate these contributions to obtain the string 
theory background for a finite 'tHooft coupling. In other words,  
spacetime perturbation theory leads to a worldsheet perturbation 
theory. 

In the next section we see how many of these  
elements are concretely realised in Kontsevich's 
matrix model derivation of Witten's conjecture. 

\newsec{The Kontsevich Model}

Kontsevich proposed a one matrix model whose free energy served as the 
generating function of the closed string correlators 
of topological gravity. We will see that his derivation of this result 
can be viewed as a special case of our general approach. Being very 
explicit it has the virtue of illustrating several features. In 
particular, he connects the flat measure over the space of Schwinger
parameters to a top form on ${\cal M}_{g,n}$ which shows how 
the closed string correlators emerge from the integrand of the Schwinger
parameter space. 

To illustrate our points we will actually reverse the logic of Kontsevich.
Let us start from the Kontsevich Hermitian matrix model 
\eqn\kont{Z(\L)=c(\L)^{-1}\int[DM]e^{-{1\over 2}Tr[\L M^2-i{M^3\over 3}]},}
where
\eqn\clam{c(\L)=\int[DM]e^{-{1\over 2}Tr(\L M^2)}.}
The constant matrix $\L$ contains the couplings of the theory via the 
dictionary
\eqn\miwa{t_i= -(2i-1)!!~Tr\L^{-(2i+1)}.}
We can choose $\L$ to be a diagonal matrix with entries 
$\L_{ab}=\L_a\d_{ab}$.
What is of interest is the free energy of this matrix model which 
is a sum over the connected vacuum diagrams. 

The vertices are all cubic and 
the propagator is given by  
\eqn\prop{<M_{ab}M_{cd}>={2\d_{ad}\d_{bc}\over \L_a+\L_b}.}
So the Feynman diagrams are a sum over all 
double line graphs $\G$ with cubic vertices which can be written as
\eqn\free{\eqalign{F(t_0,t_1,\ldots)=&\sum_{\G}{({i\over 2})^{V_{\G}}
\over |\G|}
\sum_{(r(a),r(b))}\prod_r{2\over \L_{r(a)}+\L_{r(b)}} \cr
=&\sum_{\G}{({i\over 2})^{V_{\G}}2^{E_{\G}}\over |\G|}
\sum_{(r(a),r(b))}\int\prod_r dl_r
e^{-\sum_r l_r( \L_{r(a)}+\L_{r(b)})}\cr
=&\sum_{g,n}{1\over n!}\sum_{\G_{g,n}}\sum_{a_i}
{(-1)^n 2^{2g-2+n}\over |\G_{g,n}|}\int\prod_r dl_r
e^{-\sum_i\L_{a_i}p_i}.}}
Notice that the $l_r$ that have been introduced are Schwinger parameters 
for the propagator. It will be important for the derivation that these
are later identified with the Strebel lengths. 
In the first line, the $r(a), r(b)$ denote the two 
colour indices associated with the double line 
for edge $r$. These are to be summed over,
subject to the constraint that in every closed loop a single colour flows. 
In fact, in the third line, this constraint has been explicitly taken into 
account. The sum is now only over the independent colour indices $a_i$. Where
$i$ labels the $n$ different faces of the graph. (Here we have also organised 
the graphs $\G$ by genus $g$, with $n$ marked faces $n$.) 
In the exponent we have gathered the $\L$'s with the same 
colour index and therefore it multiplies the circumference
\eqn\pa{p_i=\sum_{r_i=1}^{m_i} l_{r_i},} 
i.e. sum
over the lengths of all the $m_i$ edges that appear in the $i$'th loop. 
The symmetry factors 
of the graphs have been denoted by $|\G|$. (There is an extra symmetry 
factor of $n!$ in going to graphs $\G_{g,n}$ with $n$ marked faces.)  

At this stage we have a expressed the matrix model free energy as an
integral over the space ${\cal M}_{g,n}^{comb}$
of ribbon graphs with Schwinger parameters for each edge.  
One of the important steps in Kontsevich's derivation is the conversion
of the flat measure $\prod_r dl_r$ on ${\cal M}_{g,n}^{comb}$
to one on ${\cal M}_{g,n}\times R_+^n$. 
Kontsevich obtains the following relation
\eqn\meas{2^{5g-5+2n}\prod_{r=1}^{6g-6+3n} dl_r=\prod_{i=1}^n
dp_i\times {\O^d\over d!}.}
where the $p_i$ are the circumferences of the loops \pa\ and
parametrise the $R_+^n$. While ${\O^d\over d!}$ will be identified 
with a top form $(d=3g-3+n)$ on ${\cal M}_{g,n}$. In fact,  
\eqn\top{\O=\sum_{i=1}^np_i^2\o_i; ~~~~~~~~~~ \o_i=\sum_{1\leq r_i <r'_i<m_i-1}
d({l_{r_i}\over p_i})\wedge d({l_{r'_i}\over p_i}).}
Kontsevich identifies the $\o_i$ above with $c_1({\cal L}_i)$, the first 
Chern class of the cotangent line bundle $T^*\Sigma_{g,n}|_{z_i}$ at the 
puncture $z_i$. (See, for instance, \zvon\ for a 
detailed explanation of this identification.)
This step requires that the $l_r$ be the 
Strebel lengths i.e. the widths of the strips making up
the closed string cylinders. 

With this important step, Kontsevich has now converted the integral 
over the Schwinger parameters $l_r$ into one over 
${\cal M}_{g,n}\times R_+^n$, with the Jacobian giving a very
natural top form on ${\cal M}_{g,n}$. Integrating ${\O^d\over d!}$
over  ${\cal M}_{g,n}$ gives various intersection numbers of 
the classes $\o_i$. These are exactly the correlators of topological 
gravity according to Witten \wittop . So we have, not just any integral 
over moduli space, rather, a very natural one from the point of 
view of the closed string worldsheet. 

Kontsevich then goes on to perform the integral over the $p_i$ and thus
obtain a generating function for the intersection numbers (i.e. closed string 
correlators).
\eqn\konta{\eqalign{F(t_0,t_1,\ldots)=&
\sum_{g,n}{(-1)^n\over n!}\sum_{a_i}
\sum_{\{d_i\}}\int_0^{\infty}\prod_{i=1}^n dp_i {p_i^{2d_i}\over 2^{d_i}d_i !}
e^{-\L_{a_i}p_i}\int_{{\cal M}_{g,n}}\prod_{i=1}^n\o_i^{2d_i}\cr
=&\sum_{g,n}{(-1)^n\over n!}\sum_{a_i}
\sum_{\{d_i\}}\prod_i{(2d_i)!\over 2^{d_i}d_i!}\L_{a_i}^{-(2d_i+1)}
\int_{{\cal M}_{g,n}}\prod_{i=1}^n\o_i^{2d_i}\cr
=&\sum_{g,n}{1\over n!}\sum_{\{d_i\}}\prod_i t_{d_i}<\prod_i V_{d_i}>.}}
In the summation over $d_i$ it is understood that $\sum_{i}d_i=d$.
The last line has used the definition \miwa\ to write the final
expression in the desired form of 
generating function of closed string correlators. 
In fact, this could
be exponentiated and interpreted as a closed string theory in a background 
specified by the couplings $t_i$.

We can take away a few lessons  for our general 
approach from this derivation. 
Firstly, the identification of the Schwinger parameters 
with the Strebel lengths is crucial here. It allows the identification
of the $\o_i$ defined in \top\ with the tautological classes 
$c_1({\cal L}_i)$ on moduli 
space. Secondly, the change of measure from the flat one for Schwinger 
parameters to one on ${\cal M}_{g,n}\times R_+^n$ gives rise to a 
top form on ${\cal M}_{g,n}$. Notice that this 
change of variables is common to
all Schwinger integrals in field theory. Which 
means that this particular piece of the closed string correlators would be 
{\it common} to all the 
putative string duals to field theory. This is a nice
feature since these are the basic $2d$-gravity correlators \wittop\dist\
\dijkwit\vv\
which one
might expect are present in all closed string correlators. 

It would be nice to understand the Chern-Simons topological string 
duality \gv\ in such terms. Like the Chern-Simons theory
the Kontsevich model has also been interpreted as an open 
string field theory \gr\ and used to illustrate open-closed 
string duality. The matrix model variables have been understood in terms of 
strings stretching between FZZT branes \gr\seib . For generalisations of this 
system see \aki .

\newsec{The Four Point Function}

How can we use this approach to open-closed string duality to further
our understanding of the closed string duals to field theories? 
The dictionary \dic\ applied to field theory expressions such as 
\schem\ gives, in principle, a candidate closed string correlator on moduli 
space. We would like, for instance, to
check for a general field theory whether this is indeed a 
correlator of a consistent worldsheet CFT. The only way to do this,
at present, seems to require us to know the correlator as a function of the 
holomorphic coordinates on moduli space. In these coordinates various 
properties of the CFT would be manifest. 

To do this we should write the Strebel lengths $l_r$ in terms
of the complex coordinates $z_a$ on moduli space (and the residues $p_i$). 
This is a mathematically well posed problem. By constructing the 
unique Strebel differential on ${\cal M}_{g,n}\times R_+^n$ we obtain
the relation between the $l_r$ and the $z_a$. The problem for us is that 
the explicit construction of the Strebel differential at an arbitrary 
point in moduli space is not an easy task. As we will see, 
one can easily write down the general quadratic differential 
having double poles at $n$ specified points. This is actually 
a vector space 
of complex dimension $3g-3+2n$. Fixing the residues at the $n$
double poles cuts the dimension down by $n$. Nevertheless, we
have a $3g-3+n$ dimensional space of quadratic differentials. To  
find the unique Strebel differential in this space, 
we need to further impose the reality of various Strebel lengths \strl .
This condition is generally an implicit one for the parameters of the
quadratic differential. To solve for this condition and fix the Strebel 
differential is therefore not easy. This is what makes the task of 
expressing the $l_r$ as an explicit function of the $z_a$ difficult.
 
Let us illustrate this by looking at the simplest non-trivial case,
namely, that of the four punctured sphere. We will use  
$SL(2,C)$ invariance to place the punctures at $(1,\pm t, \infty)$.
The general quadratic differential with double poles at $(1,\pm t, \infty)$
is of the form. 
\eqn\qd{\phi(z)dz^2= -({a\over 2\pi})^2{\prod_{i=1}^4(z-z_i)dz^2\over
(z-1)^2(z^2-t^2)^2}.}
The numerator is fixed to be a polynomial of degree four because of the 
requirement of a double pole at $\infty$. 
We thus have a vector space of five complex parameters $(a,z_i)$. 
Fixing the four real residues 
$(p_1, p_{\infty}, p_{\pm})$ gives four complex conditions on the $(a,z_i)$.
Note that these are all algebraic conditions.
But we still have one complex or two real parameters which are not fixed
thus far. As mentioned above, these are fixed by demanding reality 
of two of the Strebel lengths \strl\ between the zeros $z_i$ 
\foot{Since we have already imposed the conditions 
on residues, which are sums of Strebel lengths, there are only two
independent reality conditions that we need to impose on the Strebel 
lengths. The rest are automatically real once these are imposed.}.
We see that the condition
\eqn\imstrl{Im \int_{z_i}^{z_j}\sqrt{\phi(z)}dz =0}
is a transcendental condition on the parameters $(a,z_i)$. For  
$\phi(z)$ as in \qd\ the integral in \imstrl\ is an elliptic integral and
is not easy in general to solve 
\foot{See \bel\moel\ for a combination of numerical and analytic approaches
to solve the Strebel conditions in the restricted case, arising 
in closed string field theory,
where the residues
are all equal.}. 
This is the obstruction to explicitly
obtaining the Strebel differential and thus the Strebel lengths/Schwinger
parameters
\eqn\strll{l_{ij}=\int_{z_{i}}^{z_j}\sqrt{\phi(z)}dz.}

Since the string amplitudes are integrals over the entire moduli space,
it might seem that we need the change of variables at a generic point
on the moduli space. We would thus be up against 
a technical roadblock in extracting the 
physical CFT correlators from the Schwinger parametrised expressions
of the field theory. 

But, fortunately, this is not quite true. 
As we will now explain, by going to 
appropriate kinematical limits of the spacetime correlator, we can 
focus our attention on {\it limiting regions of moduli space}.
In these limits, we will see that there are some simplifications.

The kinematical limits mentioned above are different 
UV limits in the field theory correlator. 
The dominant contribution to the Schwinger integrand 
in these limits will actually come from the region of the moduli 
space ${\cal M}_{0,4}$ where two of the punctures come together. 
In other words, we 
will really only need the relation between the Strebel lengths 
and the complex parameter $t$ (in the notation of \qd\ )
in the regime where $t\R 0$. Note that
this is precisely the region where we expect the integrand to satisfy
a {\it worldsheet} operator product expansion. Having an explicit relation
between the Schwinger parameters and $t$ in this limit will give us a 
way to check whether this is actually so. And if so, we can, in principle, 
extract out the worldsheet OPE coefficents and thus get some insight 
into the worldsheet CFT. We hope to carry this through fully later. 
Here we will only set up the limiting forms of the Strebel differential 
and show that its scaling implies the existence of a worldsheet expansion 
in powers of $\mt^{1\over 2}$.

\subsec{UV Limits of the Field Theory}

For concreteness, let us stick to a simple free field planar four point 
correlator of the form
\eqn\frpt{G^{(4)}(k_1, k_2, k_3, k_4)=
\la \Tr\Phi^{3}(k_1)\Tr\Phi^{3}(k_2)\Tr\Phi^{3}(k_3)\Tr\Phi^{3}(k_4)\ra.} 
There are a handful of graphs which contribute to the connected piece of this 
correlator. We will focus on the one which involves the 
maximal number of effective Schwinger parameters. 
This is the tetrahedral graph shown in Fig. 6. There is no 
partial gluing to be done here and therefore it is also the skeleton graph. 
As shown in the figure, it has six Schwinger parameters $\t_{ij}$,
the correct number to parametrise a open set in ${\cal M}_{0,4}\times R_+^4$. 
This is the reason we chose this correlator: it has the minimal number of 
fields which yields a tetrahedral skeleton graph. The other graphs 
that contribute to this correlator have skeleton graphs which have fewer
edges. They will be special cases (some of the $\t_{ij}\R\infty$)
of the general case considered in the tetrahedral graph. 

\fig{Tetrahedral graph for four point function with Schwinger times for
the six edges.}
{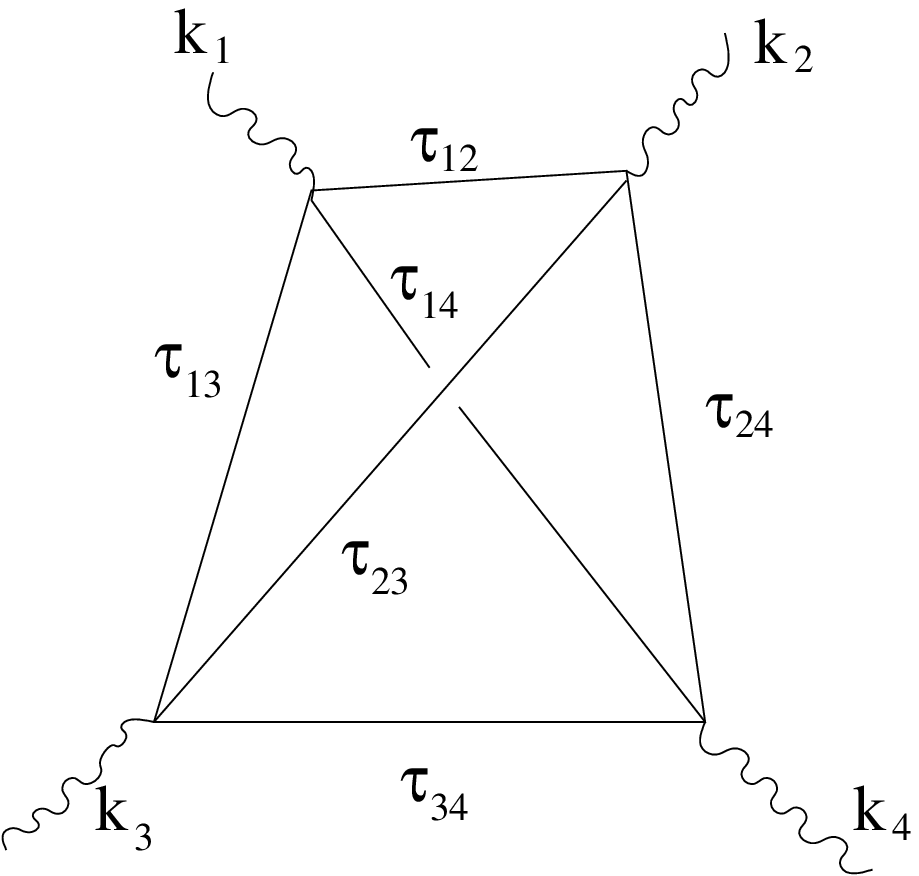}{2.5truein}

We are interested in taking the short distance or UV limit of 
this correlator.
The Schwinger parametrised form of this 
tetrahedral graph contribution to the 
four point function is
\eqn\schfr{G^{(4)}(k_1, k_2, k_3, k_4)= 
\int_0^{\infty}{\prod_{(ij)}d\t_{ij}\over \D (\t)^{d\over
2}}\exp{\{-P(\t, k)\}}.} 
We can use the explicit graph theoretic expressions for 
$\D (\t)$ and $P(\t, k)$ given in \ffb . These are written out 
for this graph in various forms in Appendix A. 
For our purpose of taking UV limits, a useful hybrid form 
is in terms of 
$\t_{12},\t_{34}$ and $\s_{ij}$ for the rest of the edges. This is 
given in Eqs. A.5-A.8. Let us first take the limit 
$k_{12}^2 \propto (k_1-k_2)^2\R \infty$, (keeping other momenta finite).
This corresponds in position space to taking the separation 
$|x_1-x_2|^2\R0$. As can be seen from Eq. A.8 the non-zero contributions
in this short distance limit  
come from a region of the Schwinger parameter space where 
$\t_{12}\s\propto {1\over k_{12}^2}$
i.e. the ratio ${\s\over \s_{12}}\R 0$. If $k_{34}$ 
is finite, then we also see from A.8 that ${\s \over \s_{34}}$ is 
finite in this limit. So we essentially have $\s_{12}\R \infty$ 
while all other Schwinger parameters are finite. This is physically 
reasonable since we expect a UV limit to correspond to  
a proper time (in this case $\t_{12}$) going to zero. 

In fact, the entire spacetime OPE expansion can be generated by 
making an appropriate expansion of the Schwinger integrand \schfr\ 
(using expressions Eq. A.6 and A.8) in powers of $\t_{12}$. We keep the 
exponential term  $e^{-\t_{12}k_{12}^2}$, but expand all other terms in 
$P(\t,\s, k)$ and $\D(\t,\s)$ containing $\t_{12}$, in a power series. 
Since 
$\t_{12}\propto {1\over k_{12}^2}$, carrying out the $\t_{12}$
integral, essentially gives more and more inverse powers of $k_{12}$.
These are the different terms in the spacetime OPE. For instance, the 
leading term in the $\t_{12}$ expansion corresponds to the integrand of 
the three point function $\la \Tr\Phi^{4}(k_1+k_2)
\Tr\Phi^{3}(k_3)\Tr\Phi^{3}(k_4)\ra$ coming from the leading connected 
contraction. The higher terms in the $\t_{12}$ expansion contain 
both descendants as well as other conformal primary operators 
(of the spacetime field theory). 
Note that the presence of terms linear in $k_{12}$ in the exponent
does not affect the expansion. They are actually necessary to reproduce
the appropriate tensor structure in an expansion about large $k_{12}$. 
So what we are seeing here is
how the Schwinger parametric representation implements the 
{\it spacetime} OPE. This is not too much of a surprise. 
The interesting thing for us is to identify this expansion in 
powers of $\t_{12}$ with a {\it worldsheet} expansion. 

Because of our identification of Schwinger parameters $\s$ with
Strebel lengths, we see that $\t_{12}\R 0$ 
corresponds to taking a particular 
limit in the space ${\cal M}_{0,4}\times R_+^4$. 
Effectively, therefore, only this limiting region of moduli space is 
relevant in this kinematic regime. What is this region?
We will see from a consideration of the Strebel differential 
that the limit where one of the Strebel 
lengths is much larger than the others corresponds to two punctures 
coming together i.e. $t\R 0$.
Actually, physical reasoning leads one to 
guess this conclusion.
After all, a UV limit in the field theory corresponds to an IR limit 
in the closed string dual, which in turn must come from a UV limit 
on the worldsheet. Thus, we expect to see the short distance structure 
of the worldsheet theory in this limit. In particular, if there is
a closed string dual, then the integrand in \schfr\ 
must exhibit a worldsheet 
OPE consistent with being a CFT. 

To extract this worldsheet OPE, 
we therefore need just find the change of variables 
from the Strebel lengths to $t$, {\it in the vicinity of} $t=0$.\foot{The 
${\cal M}_{0,4}$ 
is the part where the change of variables is non-trivial. The $R_+^4$ part is 
trivially
given by the sum of the Strebel lengths/Schwinger parameters around each 
face of the tetrahedron.} 
In other words, it is sufficent to consider a 
{\it tangent space} approximation to the moduli space ${\cal M}_{0,4}$
around the singular (stable) curve $t=0$.

For the present, we will only construct the scaling behaviour 
of the limiting Strebel differential. The change of variables following
from this construction as well as its use in obtaining a systematic 
expansion in powers of $|t|$ will be postponed for later.

Another interesting UV limit in the field theory
is when {\it both} $k_{12}^2\R 0$ and 
$k_{34}^2\R 0$. In many ways this 
is more symmetric and natural in a spacetime conformal field 
theory \foot{We thank A. Sen for this remark.}. 
In this particular limit, we see that the contributions come from
the vicinity of $\t_{12}, \t_{34}\R 0$  keeping the rest of the 
$\s_{ij}$ finite. This limit of the Strebel lengths
also corresponds to bringing the two punctures 
together, as might be expected from the previous 
physical reasoning. The difference is that the limit in the $R_+^4$ part of 
${\cal M}_{0,4}\times R_+^4$ is not the same as in the earlier case. 
In the next subsection, we will  
construct  the limiting Strebel differential in this as well as the previous 
case. 

\subsec{Limiting Strebel Differentials on ${\cal M}_{0,4}$}

Let us consider general quadratic differentials of the form \qd\ in the 
vicinity of $t=|t|e^{i\th}\R 0$. The Strebel differential is uniquely 
determined 
given $t$ and residues specified by $\{p_a\}=(p_1,p_{\infty},p_{\pm})$ at 
the double poles $(1,\infty, \pm t)$ respectively. 
Therefore, we would like to obtain $z_i=z_i(t,p_a)$ and 
$a=a(t,p_a)$ for the Strebel differential. Actually, we immediately see
from the residue at $\infty$ that 
\eqn\pinf{a=p_{\infty}.} 
This is an exact relation, valid for any $t$. 

Let us now look at the behaviour of the differential as $t\R 0$.
We will look at the two cases mentioned in the previous subsection. Namely, 
one where one of the Strebel lengths is very large compared to the others. 
Or equivalently, when all but one length is scaling to zero, with the relative
ratios of these finite in the limit\foot{Recall that we have the 
freedom to make an overall scaling of all lengths. This does not affect the 
parametrisation of the closed string surface.}. 
The other case is when two lengths
on opposite sides of the tetrahedron (the duals to the edges $(12)$ and 
$(34)$) are kept finite while others scale uniformly to zero.

In the first case, we will take the Strebel length of the edge separating 
the poles at $z=1,\infty$ finite while all others scale to zero. 
In the second, we will take the length of the edge separating the poles
$\pm t$ finite as well. 

In these limits we would like to look at the behaviour of the zeros
$z_i(t)$. 
We will often suppress the dependence on 
the $p_a$ when it does not create confusion. Let us make the scaling 
ansatz 
\eqn\zscal{z_i(t)=\mt^{\a_i}\tz_i(t)}
with $\tz_i(t)\R \tz_i$ a finite limit as $t\R 0$.

In either limiting case, since the length of the edge
separating $z=1,\infty$ is finite 
we must have $p_1=p_{\infty}$ finite in the 
limit $t\R 0$. Looking at the residue at $z=1$ then 
immediately implies that $\a_i > 0$. 
So the quadratic differential looks to leading order like 
\eqn\zlim{\phi(z)dz^2=-({a\over 2\pi})^2{(z-\mt^{\a_1}\tz_1)
(z-\mt^{\a_2}\tz_2)(z-\mt^{\a_3}\tz_3)(z-\mt^{\a_4}\tz_4)dz^2\over
(z-1)^2(z^2-t^2)^2}.}
In the limit as $t\R 0$, we will have the limiting differential taking
the simple form
\eqn\zzlim{Lim_{t\R 0} ~~\phi(z)dz^2=-({a\over 2\pi})^2{dz^2\over
(z-1)^2}.} 
In this form we see only two of the poles at $z=1, \infty$. In fact, in 
these coordinates, the coincident point of the punctures 
$(z=0)$ is a regular point.

To look at the degenerating Riemann surface more closely, we must use the 
familiar plumbing fixture construction to go to new coordinates 
$w={|t|\over z}$. This will enable us to zoom in on the behaviour 
near the colliding poles. 
Using the rule for transformation for quadratic differentials, we have 
\eqn\wqd{
\tilde{\phi}(w)dw^2=-({a\over 2\pi})^2{\tz_1\tz_2\tz_3\tz_4 \mt^{\a_1+\a_2
+\a_3+\a_4 -2}
\over e^{4i\th}}
{\prod_{i=1}^4(w-\mt^{1-\a_i}w_i)dw^2\over
w^2(w-\mt)^2(w^2-e^{-2i\th})^2}.}
Here $w_i={1\over \tz_i}$.
In the $w$ coordinates, the poles at $z=\pm t$ are now at 
finite location $w=\pm e^{i\th}$. Here it is the other set of poles, at 
$w=0,\mt$, that appear to be colliding, exactly as the plumbing fixture
is designed to exhibit. 

We can now fix the values of the $\alpha_i$
in either of the two different limiting cases. In the first case 
where all but one edge has length going to zero with all others scaling 
uniformly to zero, we can argue either from symmetry or from a more 
careful analysis of the Strebel condition, that we must have all the 
$\alpha_i=\alpha$. Furthermore, the residues $p_{\pm}$ are seen to scale as 
$p_{\pm}\propto\mt^{2\alpha -1}$. Similarly, the Strebel length between 
any two of the zeros can be seen to scale as $\mt^{\a}$ (rescale $w$
in \wqd\ by $\mt^{1-\a}$). Requiring that these two scale in the same way 
(since all these Strebel lengths are supposed to go uniformly to zero)
fixes $\a=1$. 

In the second case, we instead demand that the length of the edge separating 
$w=\pm e^{i\th}$ also remains finite as $\mt \R 0$ (while others
scale uniformly to zero). This immediately 
implies that the residues $p_{\pm}$ are finite as well 
and therefore, from  \wqd , again since
all $\a_i$ are equal, we must have $4\a-2=0$ i.e. $\a={1\over 2}$. 

Thus we can readily fix the scaling of the zeros and therefore the Strebel 
lengths in both the limits. This is interesting because we see, for instance,
in the second case that the Strebel lengths vanish as $\mt^{1\over 2}$.
At the same time, the circumferences $p_i$ are finite. 
In other words, $\t_{12}$ and $\t_{34}$ vanish as $\mt^{1\over 2}$. 
As we argued earlier, the spacetime OPE corresponds to viewing 
the Schwinger integrand in an expansion in powers of $\t_{12},\t_{34}$.
We now see that this corresponds to a worldsheet expansion in powers of 
$\mt^{1\over 2}$. This is already quite nice, since we know that the general
worldsheet OPE in a CFT can have powers of $\mt^{1\over 2}$ where $t$ is
the separation of the punctures. In fact, such a fractional power is perhaps a 
signature of a dual fermionic string theory. Of course, locality on the 
worldsheet would be a consistency condition on any
Schwinger integrand for it to qualify as a candidate string correlator. 
In other words, even though we are seeing half integral powers here, 
the sum of all contributions to an amplitude must
be single valued as a function of $t$. This 
would be quite a non-trivial check from 
the point of view of the field theory. 

In any case, in the above we have set up the limiting Strebel differentials 
for a systematic study in an expansion in powers of $\mt$. 
To be able to extract the precise worldsheet OPE, we will need to 
proceed further and get the exact dependence of the Strebel lengths 
on the $p_i$ and $\theta (={\rm arg} \ t)$, in this systematic expansion. 
We plan to return to this in later work.

\medskip
\centerline{\bf Acknowledgements}

It is a pleasure to thank D. Astefanesei,
S. Bhattacharya, J. R. David, K. Furuuchi, F. Gardiner, D. Ghoshal, 
S. Govindarajan, W. Harvey, 
S. Minwalla, A. Mukherjee, S. Mukhi, K. S. Narain, D. Surya Ramana,
K. P. Yogendran  and especially 
A. Sen for helpful discussions. I would also like to thank N. Moeller, 
M. Mulase and B. Safnuk for useful correspondence regarding Strebel 
differentials. I thank K. Furuuchi for a very careful reading of the draft. 
Finally, I must acknowledge the liberal support for this research 
from the Indian people.

\appendix{A}{Schwinger Parametrised Four Point Function} 

We will use the general graph theoretic expressions for the functions 
$\D (\t)$ and the Gaussian exponent $P(\t, k)$ in the 
Schwinger parametrisation
\eqn\meas{\D(\t)=\sum_{T_1}(\prod^{l}\t),} 
\eqn\gauss{P(\t,k)=\D(\t)^{-1}\sum_{T_2}(\prod^{l+1}\t)(\sum k)^2.}
For the definition of the 1-trees $T_1$ and 2-trees $T_2$ etc. entering into
this definition we refer the reader to \ffb\ where these are reviewed. 

For the particular case of the tetrahedral graph shown in Fig. 6, 
the complete expressions are  
\eqn\delfr{\eqalign{\D(\t)=&(\t_{13}\t_{23}\t_{24}+\t_{14}\t_{13}\t_{23}+
\t_{14}\t_{23}\t_{24}+\t_{14}\t_{13}\t_{24})
+\t_{12}(\t_{23}+\t_{24})(\t_{13}+\t_{14})\cr
+&\t_{34}(\t_{14}+\t_{24})(\t_{13}+\t_{23})
+\t_{12}\t_{34}(\t_{13}+\t_{14}+\t_{23}+\t_{24}).}}
\eqn\pfour{\eqalign{P(\t, k)=\D(\t)^{-1}&
[\t_{13}\t_{14}\t_{23}\t_{24}(k_1+k_2)^2+
\t_{12}\{\t_{14}\t_{13}(\t_{23}+\t_{24})k_1^2 + 
\t_{23}\t_{24}(\t_{13}+\t_{14})k_2^2 \}\cr
+&\t_{34}\{\t_{13}\t_{23}(\t_{14}+\t_{24})k_3^2
+\t_{14}\t_{24}(\t_{13}+\t_{23})k_4^2 \}\cr
+&\t_{12}\t_{34}\{\t_{14}\t_{13}k_1^2+
\t_{23}\t_{24}k_2^2+\t_{13}\t_{23}k_3^2 +\t_{14}\t_{24}k_4^2\cr
+&\t_{14}\t_{23}(k_1+k_3)^2+\t_{13}\t_{24}(k_1+k_4)^2 \}].}}
Here we have gathered together terms involving $\t_{12}$ and $\t_{34}$
since we will be interested in the limiting behaviour of these proper times
when we take various UV limits.

We will rewrite this in hybrid form in terms of $\t_{12}, \t_{34}$
and $\s_{ij}={1\over \t_{ij}}$ for the rest of the edges.
\eqn\phyb{\eqalign{P(\t,\s, k)=\D(\t, \s)^{-1}&
[(k_1+k_2)^2+ \t_{12}\{(\s_{23}+\s_{24})k_1^2+(\s_{13}+\s_{14})k_2^2 \}\cr
+&\t_{34}\{(\s_{14}+\s_{24})k_3^2+(\s_{13}+\s_{23})k_4^2 \}\cr
+&\t_{12}\t_{34}\{ \s_{24}\s_{23}k_1^2+\s_{14}\s_{13}k_2^2
+\s_{14}\s_{24}k_3^2+\s_{13}\s_{23}k_4^2 \cr
+&\s_{13}\s_{24}(k_1+k_3)^2
+\s_{14}\s_{23}(k_1+k_4)^2 \}].}} 
\eqn\delhyb{\eqalign{\D(\t, \s)=&(\s_{13}+\s_{14}+\s_{23}+\s_{24})
+\t_{12}(\s_{23}+\s_{24})(\s_{13}+\s_{14})
+\t_{34}(\s_{14}+\s_{24})(\s_{13}+\s_{23})\cr
+&\t_{12}\t_{34}(\s_{13}\s_{23}\s_{24}+\s_{13}\s_{23}\s_{14}
+\s_{14}\s_{23}\s_{24}+\s_{13}\s_{14}\s_{24}).}}
In terms of new momentum variables
\eqn\momnew{k_s={1\over 2}(k_1+k_2)=-{1\over 2}(k_3+k_4), ~~~
k_{12}={1\over 2}(k_1-k_2), ~~~~k_{34}={1\over 2}(k_3-k_4),}
\phyb\ becomes
\eqn\phyba{\eqalign{P(\t,\s, k)=&\D(\t, \s)^{-1}
[k_s^2+ (\t_{12}+\t_{34})\s k_s^2+\t_{12}\s k_{12}^2+\t_{34}\s k_{34}^2 \cr
+&2\t_{12}(\s_{23}+\s_{24}-\s_{13}-\s_{14})k_s\cdot k_{12}
-2\t_{34}(\s_{14}+\s_{24}-\s_{13}-\s_{23})k_s\cdot k_{34} \cr
+&\t_{12}\t_{34}\{(\s_{24}+\s_{13})(\s_{23}+\s_{14})k_s^2
+(\s_{24}+\s_{14})(\s_{23}+\s_{13})k_{12}^2 \cr 
+&(\s_{23}+\s_{24})(\s_{13}+\s_{14})k_{34}^2 
+2(\s_{13}\s_{24}-\s_{14}\s_{23})k_{12}\cdot k_{34} \cr
+&2(\s_{23}\s_{24}-\s_{13}\s_{14})k_s\cdot k_{12}+ 
2(\s_{13}\s_{23}-\s_{14}\s_{24})k_s\cdot k_{34}\}].}}
Here $\s\equiv\s_{13}+\s_{14}+\s_{23}+\s_{24}$.

Note that the dependence on the $\s$'s and the $\t$'s in this expression 
are quite suggestive. For example, many of the particular 
sums of $\s$'s which appear are given in terms of the circumferences 
$p_i$ alone of the critical graph 
of the closed string surface. We expect that this is a signature of a nice
form for the dual string correlator once it is expressed fully in terms of 
$(t, p_i)$.

\listrefs

\end